\documentclass[5p, review]{elsarticle}

\usepackage{booktabs}
\usepackage{multirow} 
\usepackage{hyperref} 
\usepackage{graphicx}
\usepackage{makecell}
\usepackage{subfigure}

\usepackage{etoolbox}
\usepackage{soul} 

\newtoggle{finalPaper}

\setstcolor{blue}

\toggletrue{finalPaper} 

\iftoggle{finalPaper} {
	\newcommand{\addtxt}[1]{#1}
	\newcommand{\change}[2]{#2}
	\newcommand{\rmvtxt}[1]{}}
{
    \soulregister\cite7 
    \soulregister\ref7
    
	\newcommand{\addtxt}[1]{\textcolor{blue}{#1}}
    \newcommand{\change}[2]{\texorpdfstring{\st{#1}\textcolor{blue}{#2}}{#2}}
    \newcommand{\rmvtxt}[1]{\texorpdfstring{\st{#1}}{#1}}
}

\newtoggle{removeItalic}
\togglefalse{removeItalic} 
\iftoggle{removeItalic} {
    \renewcommand{\textit}[1]{#1}
}

\usepackage{color}
\usepackage{colortbl}

\definecolor{gray75}{gray}{.75}

\journal{Neurocomputing}

\begin{document}

\begin{frontmatter}

\title{When Brain-Computer Interfaces Meet the Metaverse: Landscape, Demonstrator, Trends, Challenges, and Concerns}

\author[1]{Sergio L\'opez Bernal\corref{cor1}}\ead{slopez@um.es}
\author[1]{Mario Quiles P\'erez}\ead{mqp@um.es}
\author[1]{Enrique Tom\'as Mart\'inez Beltr\'an}\ead{enriquetomas@um.es}
\author[1]{Gregorio Mart\'inez P\'erez}\ead{gregorio@um.es}
\author[2]{Alberto Huertas Celdr\'an}\ead{huertas@ifi.uzh.ch}

\cortext[cor1]{Corresponding author.}

\address[1]{Department of Information and Communications Engineering, University of Murcia, 30100 Murcia, Spain}
\address[3]{Communication Systems Group CSG, Department of Informatics IfI, University of Zurich UZH, CH--8050 Zürich, Switzerland}

\begin{abstract}
The metaverse has gained tremendous popularity in recent years, allowing the interconnection of users worldwide. However, current systems in metaverse scenarios, such as virtual reality glasses, offer a partial immersive experience. In this context, Brain-Computer Interfaces (BCIs) can introduce a revolution in the metaverse, although a study of the applicability and implications of BCIs in these virtual scenarios is required. Based on a limited number of publications, this work reviews the applicability of BCIs in the metaverse, analyzing the current status of this integration based on different categories related to virtual worlds and the evolution of BCIs in these scenarios in the medium and long term. This work also proposes the design and implementation of a general framework that integrates BCIs with different data sources from sensors and actuators (e.g., VR glasses) based on a modular design to be easily extended. This manuscript also validates the framework in a demonstrator consisting of driving a car within a metaverse, using a BCI for neural data acquisition, a VR headset to provide realism, and a steering wheel and pedals. Four use cases (UCs) are selected, focusing on cognitive and emotional assessment of the driver, detection of drowsiness, and driver authentication while using the vehicle. The results demonstrate the applicability of BCIs to metaverse scenarios using the proposed framework, achieving over 80\% F1-score for all UCs, with performance close to 100\% for detecting emotions and authenticating users. Moreover, this manuscript offers an analysis of BCI trends in the metaverse, also identifying future challenges that the intersection of these technologies will face. Finally, it reviews the concerns that using BCIs in virtual world applications could generate according to different categories: accessibility, user inclusion, privacy, cybersecurity, physical safety, and ethics.
\end{abstract}


\end{frontmatter}


\section{Introduction}
\label{sec:introduction}
The term metaverse is formed by the combination of the Greek prefix \textit{Meta}, which can be translated as ``after'' or ``transcendence'', and the suffix \textit{verse}, derived from the word ``universe'' \cite{Zhang:education:2022}. A metaverse represents a persistent 3D virtual world that coexists with the physical reality, although it allows surpassing the physical limitations of the real world, such as space and time. In this environment, multiple users interact as in real life, using an avatar representing their digital alter ego. In this context, the metaverse has gained relevance in recent years, with popular implementations such as Second Life. Additionally, theoretical definitions emerged as well, like the OASIS, the metaverse introduced in the science fiction novel Ready Player One \cite{Mystakidis:metaverse:2022}.

The literature has identified several requirements to implement future metaverses characterized by complex and realistic simulations \cite{Wang:survey_security:2022, Zhang:education:2022}. First, they should ensure users' immersiveness, offering realistic scenarios that allow removing real-life limitations of space and time. In addition, they require to be economically sustainable and independent, defining a closed-loop market that generates wealth. As a third pillar, the interoperability between metaverses is essential, where virtual elements can be moved between simulated worlds. Furthermore, the scalability in these scenarios is also relevant as both technical requirements and the number of users can increase over time. Finally, a desired capability of metaverses is to be heterogeneous, supporting various technologies, protocols, data representations, and virtual spaces. 

Due to the seamless vision of the metaverse, it can be applied to many scenarios. Nowadays, metaverse solutions are most commonly used for video games, although they are also relevant in education as they introduce great versatility in teaching strategies. Nevertheless, the social component is also relevant for metaverse applications, where social relationships and marketing campaigns can be studied and developed \cite{Narin:analysis_metaverse:2021}. Additionally, training applications are deployed in the metaverse, as is the case for driving and flight simulations, due to their improved realism \cite{Njoku:metaverse_driving:2022}. Finally, other emerging applications have been considered, like telemedicine \cite{Park:metaverse_challenges:2022}. 

Although the metaverse offers evident advantages, its current maturity presents several limitations \cite{Park:metaverse_challenges:2022}. First, most metaverse applications do not manage vast volumes of users nowadays, being typically proofs of concept that cannot scale adequately. Secondly, both hardware and software available nowadays are not mature enough to offer a considerably immersive metaverse since sensations are still better felt in the physical world. In this context, the most sophisticated technologies used nowadays in the metaverse for user interaction are those considered Extended Reality (XR), which includes Virtual Reality (VR) and Augmented Reality (AR), although these technologies only intervene with visual inputs in a limited way. Furthermore, hardware accessories such as haptic gloves, eye-tracking sensors, or treadmills are also used to provide inputs to the simulation, although they cannot offer a comprehensive evaluation of the user status. Finally, development communities are scarce, essential to gain experience in metaverse implementations and thus boost their maturity and reduce deployment costs.

The previous limitations generate several open challenges that must be addressed in the metaverse. In particular for the second previous limitation, there is a need for novel technologies to offer a more immersive experience, either acting as sensors receiving data from the metaverse, as actuators sending information, or both simultaneously. In this context, many works in the literature have identified Brain-Computer Interfaces (BCIs) as the key technology to achieve a complete integration between users and the metaverse in the long term \cite{Wang:survey_security:2022, DiPietro:security_conf:2021, Wang:survey:2023}. These systems allow a bidirectional interaction with the brain, allowing both neural data acquisition and neurostimulation. Moreover, BCIs can be classified based on their invasiveness levels, where non-invasive systems require the application of electrodes on the scalp, while invasive ones need a surgical procedure to place electrodes inside the skull, either on the surface of the brain or within the brain. 

In this context, BCIs are applied to a wide variety of scenarios, being the medical one the most common, where they are used for diagnostics and for treating neurological conditions by neurostimulation \cite{Khabarova:dbsParkinson:2018}. Nevertheless, they can be employed in the metaverse for various purposes. BCIs allow controlling external objects with the mind (e.g., avatars), mental spelling, authentication with brain waves, or video games and entertainment \cite{Li:bciApplications:2015, Wang:survey:2023}. Furthermore, they are widely used for cognitive and emotional assessment and cognitive augmentation, where users can improve their mental skills, being useful for metaverse applications \cite{Papanastasiou:BCI_improvement:2022}. Besides, current literature explores the feasibility of using BCIs to allow direct communication between brains, using both neural acquisition and neurostimulation capabilities \cite{Jiang:btb_brainet:2019}.

However, despite the evolution of BCIs and the existing literature linking this topic with the metaverse, their implantation in metaverse scenarios has not been studied in depth. In this context, some works highlight that BCIs are promising systems that could be useful for interacting with the metaverse \cite{Mystakidis:metaverse:2022, Park:metaverse_challenges:2022}. They even suggest that BCIs could be used to transmit remote sensations directly to the brain \cite{DiPietro:security_conf:2021}. Finally, Wang et al. \cite{Wang:survey:2023} defined the first survey studying the applicability of BCIs to the metaverse, although focusing only on non-invasive BCIs for neural data acquisition. Nevertheless, a review of the applicability and implications of BCIs in these virtual scenarios is still needed. This lack of literature generates diverse open challenges. First, it is necessary to analyze how BCIs could contribute to the metaverse to provide a seamless and immersive experience. Moreover, it is essential to measure the performance of these systems and identify the trends and challenges that BCIs present when applied to the metaverse. Finally, there is an opportunity to identify the limitations and concerns of BCIs in the metaverse. Based on these challenges, this paper presents the following contributions\addtxt{, where the first one corresponds to the main contribution of the article, followed by subsequent contributions that aim to provide incipient steps towards the integration of BCIs into the metaverse}:

\begin{itemize}
    
    \item The study of the applicability of BCIs in the metaverse to provide an immersive experience. It analyzes the current status of this integration based on different categories related to virtual worlds and the evolution of BCIs in these scenarios in the medium and long term, considering both non-invasive and invasive BCIs. 

    \item The design and implementation of a framework able to interconnect BCIs into the metaverse, managing a wide variety of sensors and actuators. This framework comprises modules that can easily be extended to increase their functionality.
    
    \item A demonstrator to validate the framework on four well-known BCI use cases in a driving metaverse: cognitive and emotional assessment of the driver, detection of drowsiness, and driver authentication while using the vehicle. This demonstrator uses a BCI, a VR headset and a steering wheel with pedals to offer an immersive experience. All use cases obtained an F1-score superior to 80\%, highlighting the performance of identifying emotions and authentication users by binary classification, achieving close to 100\% F1-score. \addtxt{It is relevant to highlight that this work does not aim to improve the existing literature in the BCI field for each use case scenario but to provide a first attempt towards integrating BCIs into diverse metaverse application scenarios.}
    
    \item An analysis of BCI trends in the metaverse, as well as identifying future challenges that the intersection of these technologies will face. 
    
    \item A review of the concerns that the use of BCIs in virtual world applications could generate according to different categories: accessibility, user's inclusion, privacy, cybersecurity, physical safety, and ethics. 
\end{itemize}

The remainder of this paper is organized as follows. Section~\ref{sec:related} presents the related work of the metaverse and the current status of BCIs in the literature. After that, Section~\ref{sec:landscape} describes the current status, medium term and long term of the evolution of BCIs in the metaverse, analyzing both metaverse application scenarios and the integration of the human senses in virtual worlds. Moreover, Section~\ref{sec:framework} introduces the design and implementation of a framework to allow the interconnection of BCIs into the metaverse. Section~\ref{sec:demonstrator} subsequently presents the demonstrator defined to validate the framework, providing the performance of the framework on four use cases, whereas Section~\ref{sec:discussion} offers a discussion of the results. After that, Section~\ref{sec:trends_challenges} introduces the trends and challenges of BCIs in the metaverse, while Section~\ref{sec:concerns} highlights the concerns introduced by these interfaces. Finally, Section~\ref{sec:conclusions} presents the conclusions and future work.

\section{Related Work}
\label{sec:related}
This section presents the state of the art of both metaverse and BCI solutions, highlighting the most relevant works and solutions in the related work from the perspective of different application scenarios. 

\subsection{The metaverse nowadays}
One of the primary uses of the metaverse nowadays is for online multi-user video games, with examples such as Minecraft, Fortnite or World of Warcraft. Nevertheless, video games are gaining popularity for their application in other domains, such as education, for providing disruptive learning strategies based on gamification \cite{Narin:analysis_metaverse:2021}. Moreover, the literature has also explored the use of video games for soft skills in the metaverse, aiming to improve communication, socialization, teamwork and creativity \cite{Edward:SoftSkills:2022}. Finally, metaverse video games are also popular for assistance during medical treatments, such as neurorehabilitation therapy \cite{Park:metaverse_challenges:2022}.

From a social perspective, the metaverse intends to bring users closer, helping them to socialize and communicate, as highlighted by \cite{Edward:SoftSkills:2022}. One of the most representative examples is Second Life, an online virtual world where users can live an alternative life, evolving how people interact and communicate \cite{Narin:analysis_metaverse:2021}. In this context, recent research has designed and implemented many metaverse platforms, being an example the creation of a university campus prototype where users can interact as in real life, being helpful for online education \cite{Haihan:metaverse_campus:2021}. 

Education is also an essential application scenario for the metaverse. These simulated platforms can provide an immersive experience using gamification approaches. Moreover, they provide personalized content and strategies for each student, improving the monitoring and evaluation of their progress \cite{Contreras:education:2022}. Using the metaverse for education offers potential applications, such as blended learning (in-person and online) or assistance during experimental learning in the laboratory. Moreover, these approaches could help an inclusive education, helping students to be more comfortable and personalizing the experience to their needs \cite{Zhang:education:2022}. Finally, the metaverse has a huge relevance for driving and flight formation as they offer highly immersive scenarios \cite{Njoku:metaverse_driving:2022}.

The metaverse has also been explored from a marketing point of view. Nowadays, there needs to be more consensus on how marketing in the metaverse will be performed and its implications. In this context, Barrera et al. \cite{Barrera:marketing:2023} reviewed the literature to analyze the current trend of marketing applied to the metaverse, providing a research agenda to guide the evolution of this incipient field. In addition, the metaverse has also gained popularity in the tourism sector, as the metaverse could improve the engagement of tourists during the whole traveling experience, providing user immersion. Moreover, the metaverse could be a perfect substitute for those unable to travel \cite{Buhalis:tourism:2023}. 

Finally, the metaverse has also gained relevance in medicine. There is consensus among medical experts regarding the use of these virtual worlds in healthcare, being useful in multiple dimensions. First, metaverse solutions can be helpful in education scenarios, especially for recreating medical practices such as surgery without the risks of real interventions. Adopting the metaverse could also facilitate disease prevention, examination, diagnosis and rehabilitation, which is particularly promising for in-home care \cite{Yang:medicine:2022}. The use of metaverse also offers promising capabilities in emergency medicine scenarios, especially in prehospital and disaster medicine for education \cite{Wu:emergency_medical:2022}.

\subsection{Current status of BCIs}
BCIs have been typically used in medical scenarios for various purposes. First, BCIs are helpful for detecting neurological diseases such as epilepsy, typically using non-invasive techniques such as electroencephalography (EEG) \cite{Zhou:epilepsy_detection:2018}. Apart from that, one of the main uses of BCIs in medicine is for neurostimulation, as they are able to reduce the effects caused by several neurodegenerative conditions, such as Parkinson's disease \cite{Khabarova:dbsParkinson:2018}. Furthermore, BCIs are nowadays used for cognitive and emotional evaluation, used to assess the mental status of the users, as well as for cognitive augmentation, aiming to improve their mental capabilities \cite{Papanastasiou:BCI_improvement:2022}. 

These technologies can also help controlling external objects with the mind. In this sense, BCIs have an important role in helping people with disabilities, used in the literature for the control of wheelchairs and prosthetic limbs \cite{Lebedev:BrainMachineIF:2017}. Outside the medical field, BCIs allow the control of multitude of systems, such as robotic arms, mental spelling applications, multimedia systems, or the the avatar within a video game \cite{Li:bciApplications:2015}. Thus, the evolution of BCIs is moving from the traditional medical sector to other areas where BCIs aim to provide specific services. 

Authentication is another relevant scenario that the literature has widely studied. Since brain waves are unique to each person, BCIs have been employed as biometric systems in charge of detecting whether the user is legitimate or not. Moreover, and in contrast to other traditional biometric scenarios, such as fingerprints, brain signals can be updated in case of attack just by defining a new external stimuli presented to the user \cite{Xu:poc_authentication:2021}.

The literature is also exploring the use of BCIs for surfing the Internet mentally, known as brain to the Internet communication, where users can interact with a computer to search for online content \cite{DeOliveiraJunior:BtI:2018}. Furthermore, research has provided incipient results towards the direct communication between two brains, as well as the connection of multiple brains defining the concept of brainets \cite{Jiang:btb_brainet:2019}.

Finally, it is worth highlighting that the literature has started studying the applicability of BCIs into metaverse scenarios. In this sense, Wang et al. \cite{Wang:survey:2023} published the first survey in this direction, focusing on the advantages and uses of BCIs in the metaverse, although only focusing on non-invasive BCIs for data acquisition. However, invasive BCIs are gaining popularity due to promising projects, such as Neuralink \cite{Musk:neuralink:2019} and Synchron \cite{Opie:Synchron:2018}, where their final goal is to record and stimulate the brain with single-neuron resolution while reducing the risks of implanting the electrodes.

\section{Applicability of BCIs in the metaverse in the short, medium, and long term}
\label{sec:landscape}
Once key aspects of the metaverse and BCIs are introduced, it is important to provide an analysis of the current status, medium term, and long term of the evolution of BCIs in virtual worlds. Particularly, this study is performed twofold. First, it evaluates the benefits of using BCIs concerning different metaverse application scenarios. Secondly, it studies how BCIs can improve the interaction with the human senses, which are common to all metaverse application scenarios.
 
\begin{figure*}[ht]
\begin{center}
\includegraphics[width=\textwidth]{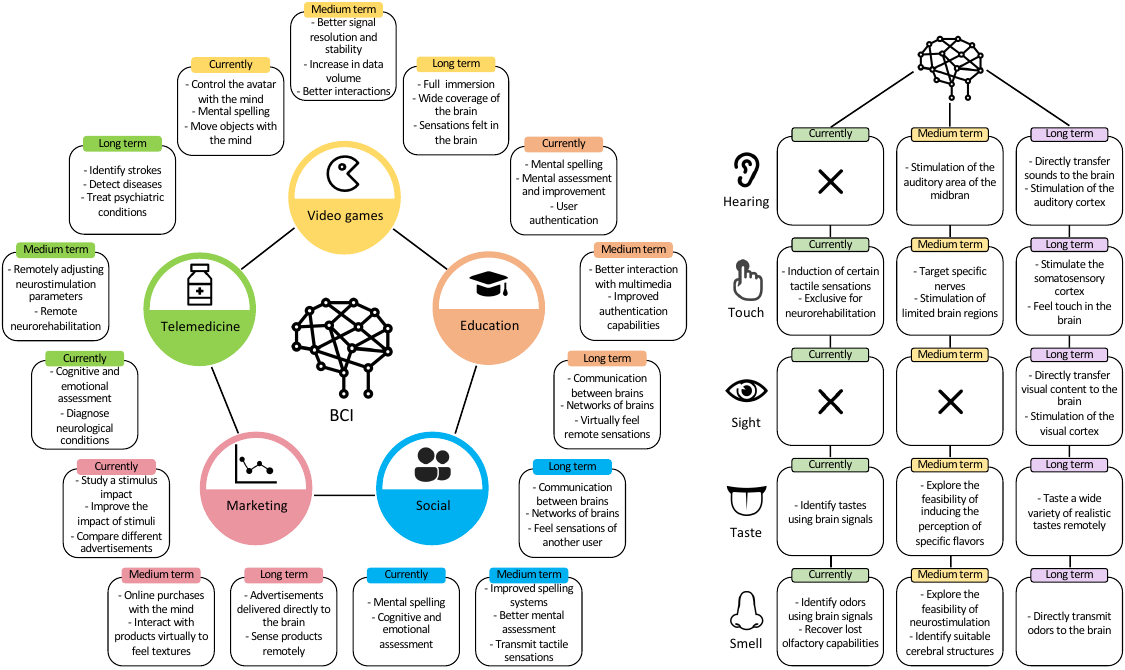}
\end{center}
\caption{Summary of the application of BCIs in the metaverse. The left side presents a study on common metaverse application scenarios. The figure on the right depicts an analysis of the human five senses.}
\label{fig:landscape}
\end{figure*}

As a first step, it is essential to clearly state the meaning of the three stages to perform a rigorous analysis. The current situation defines what is nowadays a reality and is mature in BCI research and development, although there is room for further improvement. The medium term highlights what is feasible but is under active research and presents substantial limitations. Finally, the long term highlights what is envisioned for the future and requires considerable research efforts to explore. The subsequent analysis presented for both application scenarios and human senses is summarized in \figurename~\ref{fig:landscape}.

\subsection{Metaverse application scenarios}

\subsubsection{Video games}
One of the main applications of the metaverse is in video games, where users control an avatar in different gaming scenarios \cite{Park:metaverse_challenges:2022, Narin:analysis_metaverse:2021}. However, the interaction with the avatar is typically restricted to the use of VR devices, keyboards, or mouses. In this context, BCIs based on EEG can currently provide inputs to the game, such as controlling the avatar with the mind to move in the scenario, moving objects based on limb motor imagery, or even providing mental spelling capabilities to talk during the game \cite{Kerous:games_BCI:2018}. These spellers are typically based on P300 potentials, which are physiological responses to relevant or known stimuli within a set of irrelevant ones \cite{QuilesPerez:privacy_BCI:2021}. Although the previous applications are promising, they are currently limited due to both spatial and temporal resolutions of non-invasive BCIs. Thus, BCIs cannot offer an immersive experience in gaming scenarios nowadays.

The previous limitations can be solved in the medium term using more invasive techniques, such as Electrocorticography (ECoG), offering better resolution and signal stability, although introducing physical risks \cite{Lebedev:BrainMachineIF:2017}. Moreover, an increase in invasiveness would augment data volume, offering a better interaction of users within a metaverse. In the long term, it is expected that BCIs offer a complete immersion within the game, feeling sensations realistically as they are transmitted directly to the brain \cite{DiPietro:security_conf:2021}. For that, implanted BCIs with comprehensive coverage of the brain are required since distinct brain regions manage the different human senses. The key to this evolution is to reduce the health risk introduced by invasive interfaces, which is being investigated nowadays.

\subsubsection{Education}
Another popular application of the metaverse is education, which helps provide detailed environments to understand better the contents of a topic \cite{Park:metaverse_challenges:2022, Mystakidis:metaverse:2022, Narin:analysis_metaverse:2021}. Other common uses are driving and flight training simulations, which aim to recreate a realistic experience \cite{Njoku:metaverse_driving:2022}. At the moment, the interaction with the simulation is typically reduced to VR devices, speakers or headphones, and steering wheels or joysticks for driving simulations. In this sense, BCIs can be valuable to provide spelling capabilities for writing documents or cognitive and emotional assessment to identify when users are distracted, angry, or sleepy and, thus, not in optimal conditions to do a task. Additionally, BCIs can be used for mental improvement, reinforcing users' cognition and memory \cite{Papanastasiou:BCI_improvement:2022}. Finally, BCI systems help authenticate BCI users in numerous situations, such as accessing pre-paid educational content, avoiding impersonation for cheating during exams, or controlling attendance to activities.

The advance of BCIs anticipates a revolution in the interactions in educational environments for the medium term. BCIs aim to serve as interfaces to better surf multimedia content on the Internet, used in combination with visual feedback provided by VR. Moreover, improving the temporal and spatial resolution of these interfaces can enhance their authentication capabilities due to better data quality. Finally, in the long term, BCIs intend to provide direct communication with the mind, not using microphones or speakers, which would be more efficient. In this sense, the concept of brainet, or network of connected brains, would be revolutionary for improving collaborative work \cite{Jiang:btb_brainet:2019}. Finally, users could virtually feel remote sensations in learning environments, like virtually interacting with the works displayed in a museum or experiencing a historical event firsthand.

\subsubsection{Social}
The social field is also one of the primary application scenarios of the metaverse since users also need to interact in realistic virtual worlds \cite{Park:metaverse_challenges:2022, Narin:analysis_metaverse:2021}. This interaction is currently performed using VR devices and communicating via voice or text without having further feedback from the interlocutor. Thus, BCIs can provide nowadays new ways for text inputs using mental spellers, especially relevant for the inclusion of users with disabilities (e.g., locked-in syndrome). Besides, BCIs can assess the cognitive status, which is helpful to know the engagement of the participants in a conversation or during activities in the metaverse. It is also a reality for emotional evaluation, essential for people with difficulties understanding emotions or situations \cite{Halim:poc_emotions:2020, Quiles:emotions:2023}.

Although these capabilities are helpful, it is expected that BCIs can provide in the medium term an improvement of spelling systems using more invasive technologies with better resolution, such as ECoG, as well as improving the resolution in cognitive and emotional assessment \cite{Lebedev:BrainMachineIF:2017}. The advances of BCIs may also allow the transmission of tactile sensations from the avatar to the user, improving social interactions. On its part, the long term could introduce revolutionary improvements to the metaverse, highlighting direct communication between brains, brainets for group communication, or experiencing the sensations that another person is feeling in real time \cite{Jiang:btb_brainet:2019}. These developments would be beneficial for disabled users, being capable of experiencing the entire metaverse.

\subsubsection{Marketing}
The fourth main scenario of the metaverse is marketing, which is gaining popularity due to the economic opportunities it generates in the real world \cite{Park:metaverse_challenges:2022, Narin:analysis_metaverse:2021}. Based on current capabilities in neuromarketing, BCIs are used to study the impact of a given stimulus on subjects, improve the impact of these stimuli to achieve a better result in the campaign, and compare different advertisements in the same market niche \cite{Quiles:neuromarketing:2023}. In the medium term, BCIs may bring improved integration with the metaverse to perform online purchases due to a richer resolution that would allow the acquisition of neural data with higher quality. Moreover, users could interact with the products virtually, touching them to feel their texture. In the distant future, advertisements could be delivered directly to the brain using futuristic neurostimulation BCIs, allowing remote sensing products, like seeing, hearing, smelling, or tasting them directly in our brains \cite{DiPietro:security_conf:2021}. 

\subsubsection{Telemedicine}
The last application scenario studied is telemedicine, where users can be diagnosed and treated remotely \cite{Park:metaverse_challenges:2022}. In this scenario, BCIs can contribute today to evaluating patients' cognitive and emotional status, as well as helping diagnose neurological conditions using non-invasive interfaces, as in the case of epilepsy using EEG. In the medium term, telemedicine may allow remotely adjusting neurostimulation parameters, replacing physical visits to the doctor. It could also permit remote neurorehabilitation using implantable devices, such as controlling paralyzed limbs due to damage to the nervous system \cite{Lebedev:BrainMachineIF:2017}. In the long term, BCIs might perform automatic and real-time identification of neurological strokes and emerging conditions, like recognizing epileptic seizures before their appearance. They could also be essential to treat some psychiatric conditions, avoiding the side effects of current drugs.

\subsection{The human senses in the metaverse}

\subsubsection{Hearing}
Hearing is, together with sight and touch, one of the most relevant senses to achieve an immersive simulation. Current metaverse solutions focus on using speakers or headsets, transmitting sound from the tympanic membrane to the inner ear, where the cochlea is located. Apart from these approaches, the cochlear implant is the sole technology used nowadays to transmit sound without requiring the tympanic membrane. Although these systems are safe and widely used in patients with hearing issues, they offer reduced sound quality and are unsuitable for achieving a realistic metaverse for healthy users. Nevertheless, current research explores the feasibility of using optogenetics in the cochlea for its stimulation, which can surpass current limitations. This technique genetically modifies cells to be receptive to light impulses \cite{Dieter:hearing_optogenetics:2020}. Additionally, the stimulation of the auditory area of the midbrain is under research and offers promising results \cite{Eggermont:hearing_midbrain:2017}. These approaches may define the medium term of the use of implants to improve the metaverse, although further development is needed. In this context, the final goal for the long term could be the use of neurostimulation implants in the auditory cortex in the temporal lobe, directly transmitting the information from the metaverse to the brain. Examples of promising under-research BCI solutions intending a broad coverage of the brain are Neuralink \cite{Musk:neuralink:2019} and Synchron \cite{Opie:Synchron:2018}.

\subsubsection{Touch}
Concerning touch, current metaverse solutions rely on haptic devices that transmit determined sensations to the user, typically to the hands \cite{Park:metaverse_challenges:2022}. Nevertheless, these technologies do not provide an immersive experience. Considering BCIs, current neurorehabilitation research has considerably advanced in recent years, highlighting the recovery of certain tactile sensations in patients with complete spinal cord injury by providing feedback in another part of the body \cite{Ganzer:touch_BCI:2020}. Research has also allowed feeling sensations coming from a robotic arm by using stimulation of the somatosensory cortex \cite{Flesher:touch_BCI:2021}. Nonetheless, these technologies and procedures are still under research and do not represent commercial products. In the medium term, the advances of these technologies and the expansion of implantable BCIs to the general public may induce certain tactile sensations while using the metaverse, based on targeting specific peripheral nerves or even stimulating limited brain regions. In the long run, implanted BCIs with comprehensive brain coverage could allow targeting the somatosensory cortex to recreate sensations comparable to real-life experiences. 

\subsubsection{Sight}
Sight is one of the most important senses for the metaverse since it allows the visualization of the simulated environment. Nowadays, the metaverse is explored using traditional computer screens, VR, or AR devices. Despite the advantages of these technologies, they are far from offering a completely immersive experience. This ideal experience can be provided by BCIs, helping transmit the images directly to the body without the need for external displays. Currently, the best and more secure way to directly provide images using implantable devices is visual prostheses, where retinal implants are the most relevant and safe. However, research on blind patients has just provided a certain degree of light perception, motion detection, and reading, being still immature for the metaverse as the drawbacks are greater than the benefits \cite{Niketeghad:sight:2019}. In this context, the medium term considers advancing these procedures and technologies to provide a better visual experience, surpassing current limitations regarding image quality. Finally, current advancements in visual cortex stimulation help the inclusion of invasive BCIs with broad coverage of the visual cortex to directly transfer multimedia content to the brain, although these techniques are still immature today \cite{Niketeghad:sight:2019}. It is worth noting that visual implants are not considered BCIs and, thus, the use of BCIs is only expected for the long term.

\subsubsection{Taste}
Taste and smell are two of the senses that require significant research to be included in the metaverse. Focusing on the former, although multiple proposals in the literature can be used as alternatives to interact with external devices to perceive taste, such as Taste the TV \cite{Miyashita:taste_TV:2021}, this field is still in an incipient stage. Regarding BCIs, the literature has focused on identifying tastes based on brain signals, although various factors affect the identification, such as age, gender, state of mind, or stimulus type \cite{Anbarasan:taste_BCI:2022}. Despite these advances, literature inducing taste using neurostimulation is lacking. Based on that, the medium term can include exploring and evaluating the feasibility of using BCIs to induce the perception of particular flavors. In the long term, using miniaturized electrodes could allow the generalization of BCIs, eliciting the sensation of a wide variety of realistic tastes. 

\subsubsection{Smell}
Finally, smell integration into the metaverse is limited. Current research presents prototypes to study the feasibility of these systems, which typically consist in devices attached to VR glassed to provide odors \cite{Guimaraes:olfactory_vr:2022}. Regarding BCIs, the literature is scarce and focuses on two main directions. The first consists in identifying odors based on brain signals \cite{Yang:olfactory_BCI:2022}, while the second one intends to recover olfactory capabilities in patients with a total loss of the sense \cite{Holbrook:smell_BCI:2020}. Despite the advances of BCIs regarding neurostimulation, the works addressing the evocation of odors in the brain are scarce. Based on that, there is an opportunity to continue exploring the feasibility of applying neurostimulation strategies and identifying the most suitable cerebral structures in the medium term. Based on these achievements, the long term could focus on providing broad coverage of these structures, directly sending odors to the brain.

\section{Design and implementation of the proposed framework}
\label{sec:framework}

\begin{figure}[ht]
\begin{center}
\includegraphics[width=\columnwidth]{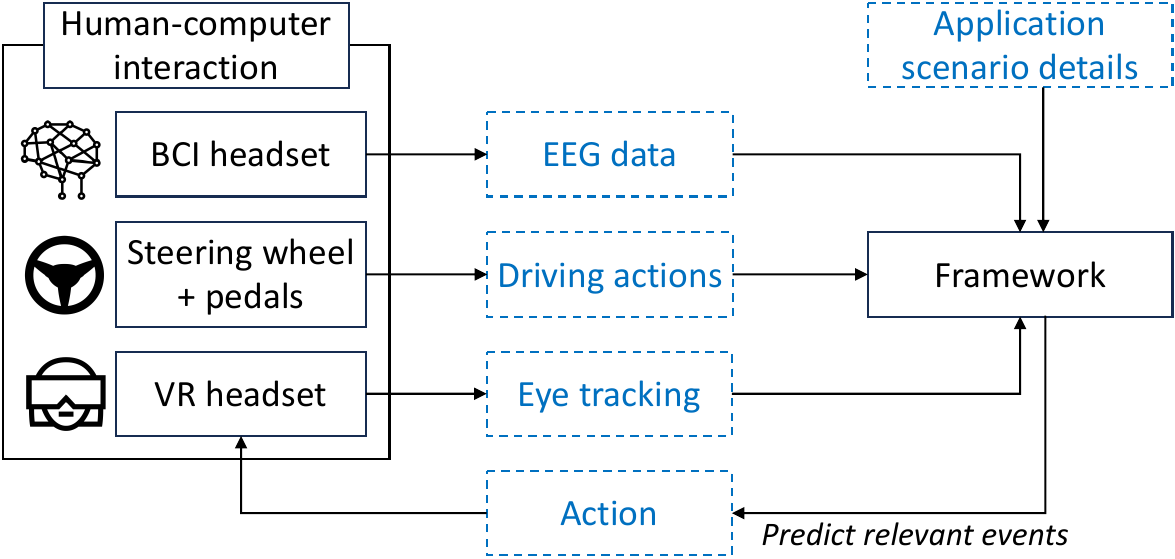}
\end{center}
\caption{General representation of the framework functionality. Blue dashed boxes indicate relevant data exchanged, either representing inputs to the framework or the outputs from the framework used as feedback to the VR headset.}
\label{fig:framework-design}
\end{figure}

This section describes relevant design and implementation aspects of the proposed framework. \change{First of all, Figure 2 presents a simplification of the functionality provided by the framework, which}{This framework} is general enough to allow the interconnection of a wide variety of heterogeneous sensors and actuators with metaverse scenarios. In particular, \change{this figure}{\figurename~\ref{fig:framework-design}} depicts examples of data reception coming from a BCI headset, a steering wheel with pedals, and a VR headset\change{as a particular setup}{, illustrating the setup implemented, which will be presented in detail later in this section}. These data are collected by the framework, which, together with information regarding the specific metaverse application scenario simulated, performs intelligent tasks to predict relevant scenario-related events. These events are then transformed into particular actions \rmvtxt{related to the application scenario} and transmitted to actuators to provide a dynamic experience, such as presenting visual events in a VR headset.

\begin{figure}[!ht]
\begin{center}
\includegraphics[width=0.9\columnwidth]{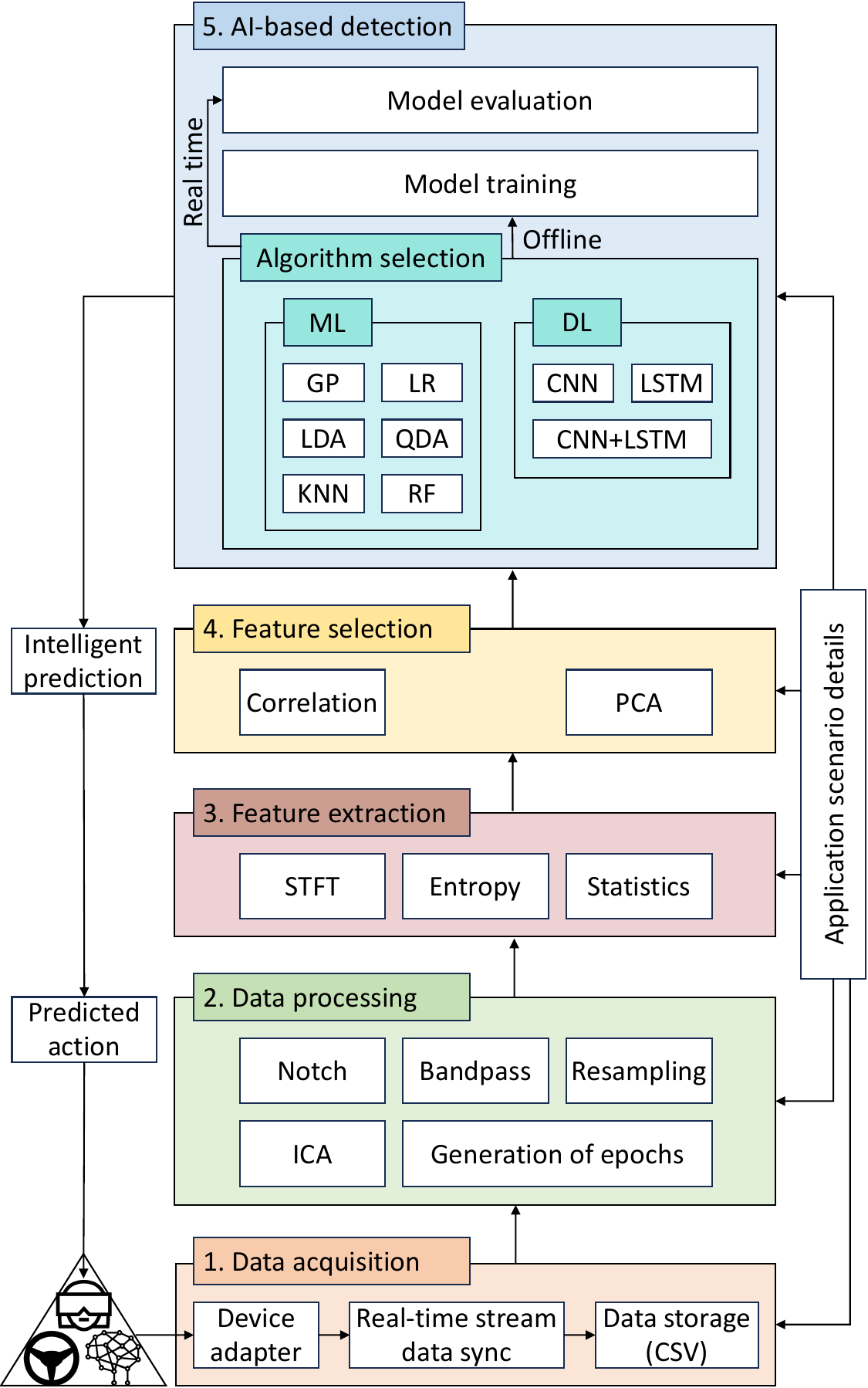}
\end{center}
\caption{Modules of the proposed framework, representing the main techniques applicable for each module implemented and the relationships between them.}
\label{fig:framework-implementation}
\end{figure}

After presenting an overall view of the external interactions of the framework, \figurename~\ref{fig:framework-implementation} depicts the main modules of the framework and the interactions between them. As observed, the modules are numbered bottom-up, highlighting their execution order. Moreover, it is relevant to note that all modules have access to the details of the application scenario, which guides the actions performed in each stage. Subsequent sections present the implementation details of each module composing the framework. 

\subsection{Data acquisition}
\label{subsec:framework-acquisition}

The first module of the framework focuses on \change{the acquisition of data}{data acquisition} from external sensors, such as BCIs or \change{driving}{steering} wheels. However, it has been conceived to be easily extended and compatible with many devices. To this end, the framework has a Device adapter component implementing the adapter \rmvtxt{pattern} software \addtxt{pattern} to easily implement the communication with external devices based on data streams. For example, for the communication with BCIs, the framework includes the Lab Streaming Layer (LSL) protocol using Python \cite{pylsl}, which is a de facto standard in the neurotechnology industry that allows the reception of brainwaves variations for each EEG channel monitored. 

After that, the data streams received in real time from different devices must be synced, incorporating a timestamp that will be useful in subsequent modules. This process allows a data fusion process that becomes essential in metaverse scenarios, where device interconnection becomes paramount. After this gathering process, the data are stored in CSV files for \addtxt{their} use by the Data processing module, employing the Pandas library from Python \cite{Pandas}. 

Finally, it is worth highlighting the importance of the details provided to the framework regarding the application scenario, which, in the specific case of the Data acquisition module, determines which devices and data representations will be used and how the synchronization of the data streams must be performed. These data inputs provide flexibility to the framework and ensure its compatibility with many simulated environments. These scenario-related details are expressed in YAML configuration files that all the modules understand. 

\subsection{Data processing}
\label{subsec:framework-processing}

This module obtains the data stored by the acquisition process and applies data processing techniques to improve data quality, such as reducing signal noise. Specifically for BCIs, this module includes several components corresponding to data processing techniques using the MNE library from Python \cite{MNE}. First, the notch filter can attenuate the interference introduced into EEG signals by the electrical cabling of the BCI. Specifically, it is commonly applied at 50 Hz in Europe and 60 Hz in the United States due to their respective electrical specifications. 

The bandpass filter can isolate a specific interval of relevant frequencies, thus discarding those over and under that range. This filter is commonly used in BCI solutions to isolate frequencies where relevant events can be found. Moreover, the resampling method can change the number of samples per second available. It is common in BCI research to use a downsampling approach to reduce the amount of data available, thus improving the speed of AI-based models both in the training and evaluation stages. Independent Components Analysis (ICA) is widely employed to isolate and discard artifacts contained in EEG signals, such as eye blinks.

Finally, after applying data processing techniques, this module can divide the data into epochs of a certain duration to be compatible with the next stages in the framework, such as feature extraction and AI-based classification. \addtxt{In the context of BCIs, an epoch corresponds to a specific, time-locked segment of recorded brain activity data}.  It is worth noting that the application of these techniques and their order will depend on the application scenario in which the framework is being utilized. Furthermore, although this framework mainly implements processing techniques for EEG data, it can be extended to include new processing mechanisms for other stream data sources. 

\subsection{Feature extraction}
\label{subsec:framework-extraction}

Once the acquired data are processed to improve \change{its}{their} quality, extracting features that better characterize the behavior of the data received is common. Particularly for the BCI domain, the Short-Term Fourier Transform (STFT) is relevant to transform EEG signals from a temporal domain to the frequency domain, enabling the analysis of relevant features, such as the Power Spectral Density (PSD). Moreover, entropy helps determine the uncertainty or randomness of the signals. Finally, many statistics can be obtained from raw EEG signals, such as the mean, maximum, and minimum voltages contained in each epoch. These features are widely used in BCI research to represent the variations and particularities of brainwaves better. \change{This functionality has been included using the}{All the previous functionality has been implemented relying on the} MNE Python library \cite{MNE}. Nevertheless, according to the application scenario specification, the framework is open for its extension to include more feature extraction techniques. 

\subsection{Feature selection}
\label{subsec:framework-selection}

The number of such characteristics can be considerable after obtaining relevant features representing the original data obtained in previous modules. This size can be a problem in constrained scenarios since AI-based algorithms will train models considering all these features, deriving in a higher computational and temporal cost. To reduce this load, multiple techniques can analyze and reduce the number of characteristics used. One of these techniques consists \change{of}{in} studying the correlation between features, simplifying those with a high correlation, implying that they are too similar. Moreover, Principal Component Analysis (PCA) can reduce the dimensionality of the features, automatically creating a lower number of features that essentially represent most of the aspects described by the original characteristics. 

\subsection{AI-based detection}
\label{subsec:framework-AI}

The last module of the framework implements a supervised AI approach to predict relevant events in the data obtained from external devices. As shown in \figurename~\ref{fig:framework-implementation}, the framework includes both ML and DL algorithms, using either raw data obtained directly from the devices or features extracted from them. Thus, the first step performed by this module is to consult the details of the application scenario to determine which set of algorithms will be employed. If no previous models are trained for a particular application scenario, the AI-based detection module initiates an offline model training process based on the previously indicated specifications. After that, this module will be capable of evaluating the models in real time without executing the Model training component, thus reducing its resource consumption and temporal requirements. 

For each new piece of information obtained by the framework, the models available will offer a real-time prediction containing specific actions to be transmitted to external devices. For example, the framework could transmit relevant visual information to the user using a VR headset, such as alerts. In the same direction, futuristic BCIs could deliver these events directly to the brain. This module has been implemented using the Scikit-learn \cite{scikit-learn}, TensorFlow \cite{TensorFlow}, and Keras \cite{Keras} Python libraries for AI. 

\section{Demonstrator of the applicability of BCI systems in the metaverse} \label{sec:demonstrator}

This section offers a demonstrator highlighting the performance of non-invasive BCIs in immersive metaverse scenarios, employing the framework presented in Section~\ref{sec:framework}. Since using metaverses for educational purposes is gaining tremendous popularity, this demonstrator focuses on a driving simulation environment with a learning approach. Nevertheless, this metaverse is not exclusive to teaching users how to drive, as it is also helpful for car companies to validate the functionality, acceptance, and performance of BCIs to be incorporated in new prototypes. More in detail, this demonstrator validates the utilization of BCIs on four use cases (UCs) of interest in BCI research, selected according to previous literature highlighting the benefits of using BCIs in simulated driving environments \cite{MartinezBeltran:safecar_distractions:2022, Zhang:poc_drowsiness:2020, Xu:poc_authentication:2021, Quiles:emotions:2023}. A description of the four UCs defined is subsequently presented:  

\begin{itemize}
    \item UC1 uses a BCI to assess the user's cognitive status to identify distractions while driving. This evaluation is essential since many accidents are caused by insufficient attention on the road, leading to catastrophic consequences.

    \item UC2 consists in detecting the emotional status of the driver using a BCI. Emotions are a paramount risk factor while driving since an angry driver tends to make mistakes or unexpected actions, dramatically interfering with driving performance.
    
    \item UC3 also evaluates the driver's cognitive status, but this time for identifying drowsiness at the wheel. In addition to distractions, drowsiness is one of the major causes of road accidents, as drivers tend to be overconfident with their driving abilities when tired.
    
    \item UC4 focuses on authenticating the user while driving using brain waves. This functionality is useful for the metaverse since it provides a seamless experience and guarantees that only legitimate users will access the simulation.
\end{itemize}

To deploy and validate the previous use cases, the present research used the DReyeVR driving simulator as a flexible and open-source solution based on Carla, a previously existing simulator \cite{Dosovitskiy:CARLA:2017}. In particular, DReyeVR extends Carla by providing virtual reality support and eye-tracking integration, relying on the existing Carla Python API \cite{Silvera:DReyeVR:2022}. 

Moreover, this work decided to use the OpenBCI EEG electrode cap in combination with the Cyton eight-channel biosensing board to acquire EEG signals since this BCI has a reduced cost, represents an open source initiative, and provides a sufficient acquisition resolution for the requirements of the present research, acquiring EEG signals with a resolution of 24 bits \addtxt{and} a sampling rate of 250 Hz. This BCI can be extended to 16 channels using the Daisy adapter \cite{OpenBCI}. Moreover, the OpenBCI cap can be easily combined with the Varjo Aero VR headset to provide an immersive visual experience while recording EEG data. In particular, this VR device provides a resolution of 2880 x 2720 pixels per eye, with a brightness of 150 nits and a refresh rate of 90 Hz. It allows a horizontal visual field of 115 degrees and supports an audio jack connection for audio using in-ear headphones or external speakers \cite{VarjoAero}. 

\begin{figure}[ht]
\begin{center}
\includegraphics[width=0.7\columnwidth]{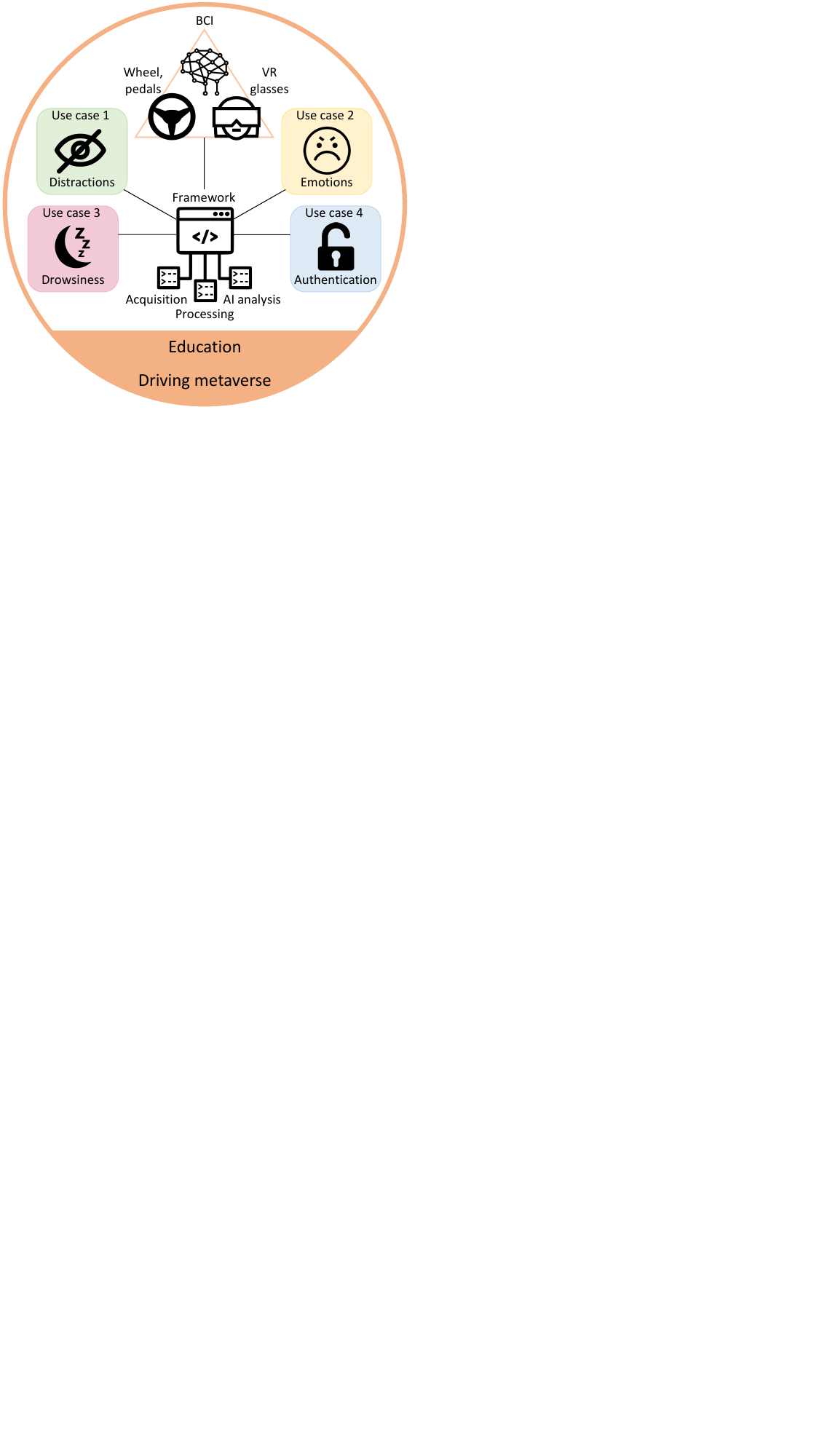}
\end{center}
\caption{Setup used to evaluate the performance of the framework in the metaverse for each use case defined.}
\label{fig:setup}
\end{figure}

This work also used a steering wheel and pedals to provide a more immersive driving experience. In particular, the experiments were performed using the KROM K-Wheel controller as it is a versatile device compatible with many open-source libraries, which allows the customization of its controls. Besides, it provides two pedals for accelerating and braking, offering a more realistic and immersive experience \cite{KROM}. In a nutshell, \figurename~\ref{fig:setup} presents the setup employed to evaluate the performance \change{in the use of BCIs}{of using BCIs} in a driving metaverse scenario, highlighting the central role of the framework developed to coordinate the actions required per UC. Subsequent sections present specific considerations for the implementation of each UC. 

\subsection{UC1: Assessment of cognitive status}
\label{subsec:demonstrator-UC1}

The first use case presents periodic visual distractions through the VR headset on top of the driving simulation to evaluate if drivers tend to get distracted. For that, the study identifies such distractions over the EEG signals acquired. More particularly, two categories of distractions are presented. First, visual distractions, such as geometric shapes, force drivers to look at the new objects randomly displayed on the screen each time. Secondly, mathematical operations intend to cause a higher cognitive load. This UC has been designed considering the work published by Martínez-Beltrán et al. \cite{MartinezBeltran:safecar_distractions:2022}. 

The acquisition process for this UC consists in acquiring EEG signals from the Fp1, Fp2, Fz, P3, Pz, P4, O1, and O2 brain locations, following the standard 10-20 system \citep{Klem:10-20system:1999}. Ten subjects participated in the experiment, six males and four females, which drove uninterruptedly for twenty minutes, alternating periods with and without distractions. Once the EEG signals were obtained, the Data processing module applied a notch filter at 50 Hz \addtxt{to remove the interference introduced by the electric system}, a band-pass filter between 0.1-60 Hz \addtxt{to select the frequencies of interest for this UC}, downsampling to 32 Hz\addtxt{, which offers an interesting trade-off between data-loss and performance for subsequent stages}, and ICA \addtxt{to remove artifacts in the EEG signals}. After this data processing stage, the signals were split into one-second epochs. 

After the processing phase, the framework obtained the PSD of the EEG signals applying STFT. Moreover, the experiments obtained the entropy of the EEG signals and several relevant statistics, such as its mean, quartiles, kurtosis, and Hjorth parameters, among others. Due to the high number of features obtained, it was necessary to apply PCA to reduce the dimensionality of the data. 

Finally, the AI-based detection module applied two supervised classification approaches. First, a binary classification intended to identify whether drivers suffered a distraction, while a multiclass classification intended to identify the kind of distraction presented to the user based on the analysis of the EEG signals captured, detecting if drivers are not distracted. For both approaches, the framework trained Support Vector Machine (SVM), k-Nearest Neighbors (KNN), and Random Forest (RF) ML-based models, studying the F1-score obtained for each subject. \addtxt{Furthermore, this UC has used StratifiedKFold for cross-validation with ten splits, selecting 70\% of the data for training and the remaining 30\% for testing. Finally, a hyperparameter search using RandomizedSearchCV has been used to identify the best configurations for each ML algorithm.}

\begin{figure}[ht]
\begin{center}
\subfigure[Binary classification assessing if drivers are distracted or not according to the F1-score performance metric.]{\includegraphics[width=\columnwidth]{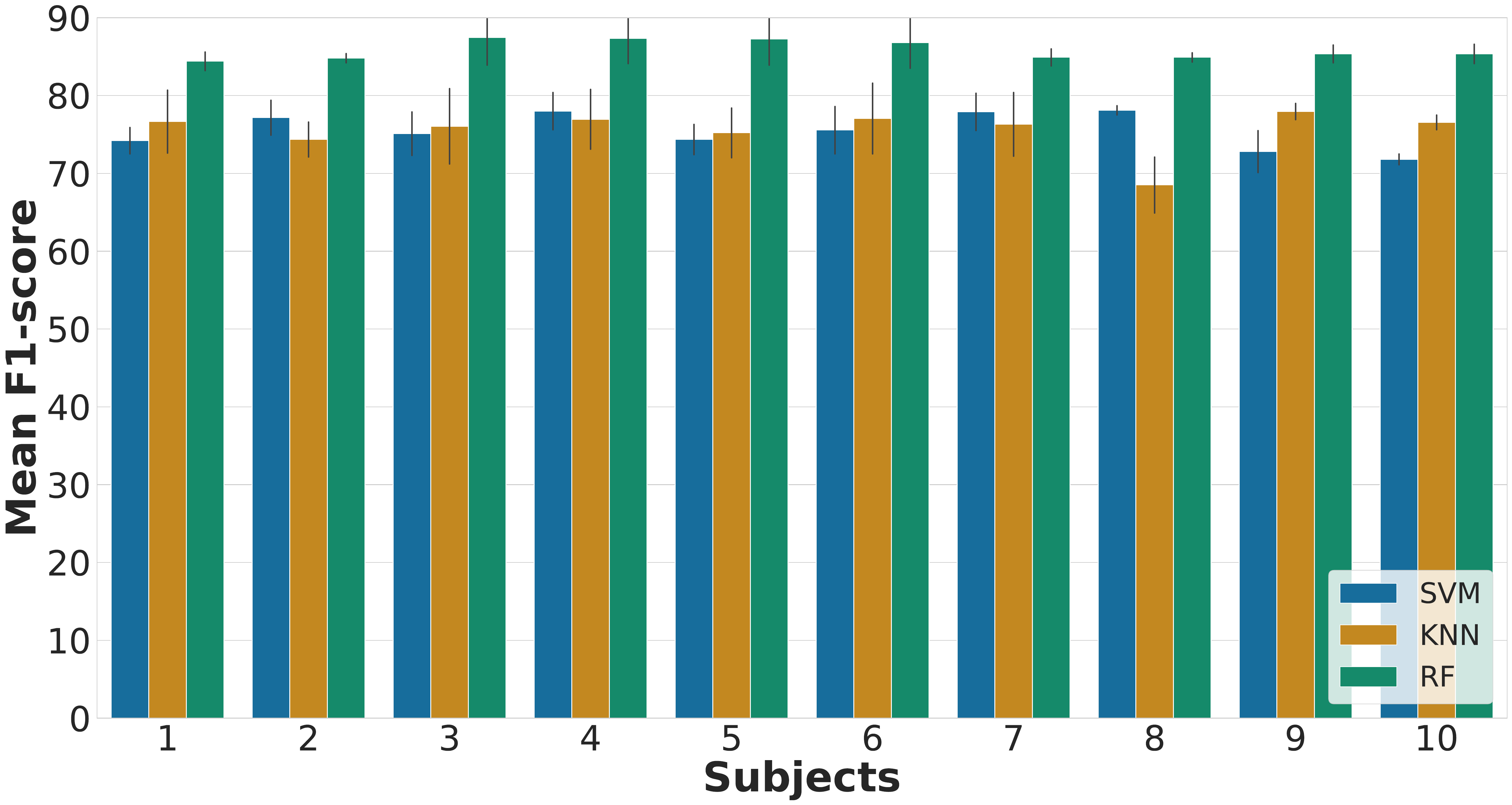}}
\subfigure[Multiclass classification determining the kind of distraction presented to the drivers.]{\includegraphics[width=\columnwidth]{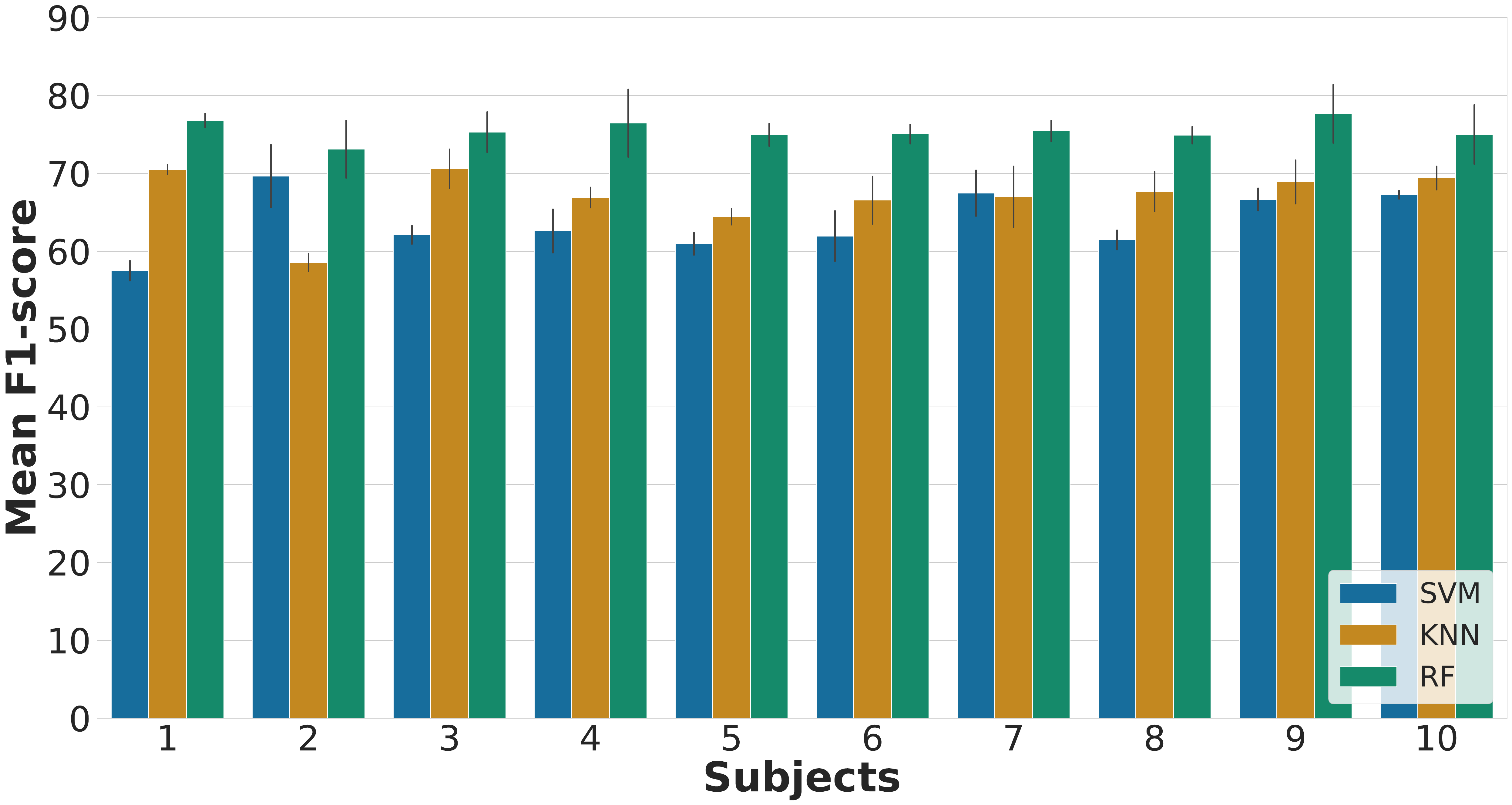}}
\end{center}
\caption{Performance obtained by the framework for UC1, considering binary and multiclass approaches to predict distractions while driving.}
\label{fig:results-UC1}
\end{figure}

\figurename~\ref{fig:results-UC1} presents the results obtained for UC1 based on mean percentage of F1-score per subject. Focusing first on binary classification, RF is the algorithm offering the best performance, over 80\% for all individual subjects tested. In addition, the multiclass classification offered excellent results, being capable of identifying the correct class of the total of three possible with more than 70\% F1-score for all subjects. 

\subsection{UC2: Identification of emotions}
\label{subsec:demonstrator-UC2}

The second use case aims to study the feasibility of detecting different emotions from EEG signals while driving. For that, the framework presents auditory stimuli to the driver to induce angry, neutral, or joyful emotions. Neutral emotions refer to mild reactions when facing external stimuli, such as nature sounds. Moreover, the lack of external stimuli is also intended to be detected, aiming to identify when the driver is not facing external stimuli. 

The methodology consists of freely driving in the simulation, alternating between emotions every five minutes. Thus, the experiment commences without any stimuli, where the subjects drive. After that, neutral stimuli are presented, such as rain sounds. Then, joy stimuli consisted of optimistic songs to cheer up the driver. Finally, heavy traffic and drilling sounds were reproduced, aiming to induce an angry mood. This sequence of emotions was repeated a total of two times to offer data variability. This methodology has been based on the previous considerations published by Quiles-Pérez et al. \cite{Quiles:emotions:2023}.

The BCI acquired data from eight electrode placements (Fp1, Fp2, F3, F4, F7, F8, Fz, and Cz) on \change{three subjects}{}: \change{two}{four} men and \change{one woman}{three women}. Apart from the BCI, it was necessary to use external speakers connected to the VR headset to reproduce the stimuli presented by the framework to the drivers. This work considered using external speakers and not in-ear headphones, as the latter option could cause interference in the EEG acquisition process. 

After acquiring the EEG data, the experimentation relied on the Data processing module of the framework to execute a notch filter at 50 Hz, a bandpass filter between 4-60 Hz, and ICA to remove artifacts. \addtxt{As a final step, the data processing stage used a four-second sliding window with a stride of one second to generate windows used by the next component}. Then, the Feature extraction module used \rmvtxt{the} STFT to extract brainwave frequency bands (theta, alpha, beta, and gamma) \addtxt{from the previous windows}. After that, this work obtained statistics from each band, such as mean and standard deviation. The feature selection process used correlation and PCA approaches to reduce the complexity of the data and thus ease further data classification tasks. 

Finally, UC2 trained ML and DL models using the AI-based detection module to classify emotions. First, a binary classification intended to differentiate between two emotions (no stimuli and angry). After that, a multiclass classification identified an emotion within three possible emotions (no stimuli, angry, and neutral). Finally, this multiclass classification was extended to cover all four emotions, thus including the joyful state of the experimentation. The ML-based algorithms tested were RF, KNN, and XGBoost\addtxt{, using hyperparameter search to find the best configuration of the models}. In contrast, Long Short-Term Memory (LSTM), Convolutional Neural Networks (CNN), and LSTM+CNN were the DL-based models tested. \addtxt{LSTM+CNN was considered to take advantage of the good behavior of LSTM when dealing with long-term relationships in the data, while CNN can extract relevant spatial features from the data.} 

\addtxt{Regarding DL configurations, all models used two layers with 32 neurons per layer, having ReLU as the activation function. Moreover, all models used a Dropout at 20\% between both layers. The main difference in terms of architecture resides in the output layer, where the LSTM and CNN models flattened the results. At the same time, the LSTM+CNN model used a perceptron layer, employing a sigmoid for binary classification and a softmax activation for multiclass emotion identification. Finally, all DL algorithms used EarlyStop to avoid overfitting.}

\tablename~\ref{tab:results-UC2} presents the results obtained for UC2, where it can be seen that the classification between two emotions generated an F1-score above \change{96}{93}\% for all subjects. Moreover, the identification of three different emotions offered results over \change{92}{90}\%. Finally, detecting all four emotions reached an F1-score over 78\% for all subjects, which represents good results considering that a random classification would correspond to an F1-score of 25\% when having four classes. It is worth highlighting that the algorithm having the best results was RF, although most algorithms provided good results for binary classification, especially those based on neural networks. However, increasing the number of emotions reduced the performance in DL-based models due to the limited data available and the large similarities between classes (see more details in \cite{Quiles:emotions:2023}).

\begin{table}[!htb]
\centering
\caption{Results for UC2, studying the performance of identifying different numbers of emotions while driving in an immersive metaverse scenario.}
\label{tab:results-UC2}
\resizebox{\columnwidth}{!}{
\begin{tabular}{@{}cccccccc@{}}
\toprule
\textbf{Experiments} & \textbf{Subjects} & \textbf{RF} & \textbf{KNN} & \textbf{XGBoost} & \textbf{LSTM} & \textbf{CNN} & \textbf{CNN+LSTM} \\ \midrule

\multirow{7}{*}{Two emotions} & Subject 1 & \cellcolor{gray75}\makecell{99.0\\\addtxt{$\pm$0.12}} & \makecell{76.0\\\addtxt{$\pm$1.02}} & \makecell{79.0\\\addtxt{$\pm$0.87}} & \makecell{90.0\\\addtxt{$\pm$0.95}} & \makecell{70.5\\\addtxt{$\pm$1.15}} & \makecell{92.0\\\addtxt{$\pm$0.75}} \\ \cline{2-8}
 & Subject 2 & \cellcolor{gray75}\makecell{96.0\\\addtxt{$\pm$0.11}} & \makecell{91.5\\\addtxt{$\pm$0.78}} & \makecell{95.0\\\addtxt{$\pm$0.85}} & \cellcolor{gray75}\makecell{96.0\\\addtxt{$\pm$0.90}} & \makecell{89.5\\\addtxt{$\pm$0.73}} & \makecell{95.0\\\addtxt{$\pm$0.65}} \\ \cline{2-8}
 & Subject 3 & \cellcolor{gray75}\makecell{97.0\\\addtxt{$\pm$0.13}} & \makecell{84.0\\\addtxt{$\pm$1.10}} & \makecell{88.0\\\addtxt{$\pm$0.96}} & \makecell{96.0\\\addtxt{$\pm$0.88}} & \makecell{89.5\\\addtxt{$\pm$0.70}} & \makecell{95.0\\\addtxt{$\pm$0.80}} \\ \cline{2-8}
 & Subject 4 & \cellcolor{gray75}\makecell{96.4\\\addtxt{$\pm$0.10}} & \makecell{75.0\\\addtxt{$\pm$1.20}} & \makecell{83.3\\\addtxt{$\pm$1.00}} & \makecell{94.6\\\addtxt{$\pm$0.85}} & \makecell{68.8\\\addtxt{$\pm$1.25}} & \makecell{95.0\\\addtxt{$\pm$0.77}} \\ \cline{2-8}
 & Subject 5 & \makecell{93.2\\\addtxt{$\pm$0.12}} & \makecell{88.2\\\addtxt{$\pm$0.90}} & \cellcolor{gray75}\makecell{96.5\\\addtxt{$\pm$0.85}} & \makecell{94.2\\\addtxt{$\pm$0.70}} & \makecell{86.6\\\addtxt{$\pm$0.95}} & \makecell{94.1\\\addtxt{$\pm$0.88}} \\ \cline{2-8}
 & Subject 6 & \cellcolor{gray75}\makecell{95.5\\\addtxt{$\pm$0.14}} & \makecell{83.6\\\addtxt{$\pm$1.10}} & \makecell{88.4\\\addtxt{$\pm$0.94}} & \makecell{91.4\\\addtxt{$\pm$0.83}} & \makecell{87.3\\\addtxt{$\pm$0.91}} & \makecell{90.3\\\addtxt{$\pm$0.87}} \\ \cline{2-8}
 & Subject 7 & \cellcolor{gray75}\makecell{97.1\\\addtxt{$\pm$0.10}} & \makecell{78.2\\\addtxt{$\pm$1.15}} & \makecell{82.1\\\addtxt{$\pm$0.98}} & \makecell{90.7\\\addtxt{$\pm$0.80}} & \makecell{71.1\\\addtxt{$\pm$1.22}} & \makecell{90.6\\\addtxt{$\pm$0.75}} \\ \midrule

\multirow{7}{*}{Three emotions} & Subject 1 & \cellcolor{gray75}\makecell{94.1\\\addtxt{$\pm$0.18}} & \makecell{60.3\\\addtxt{$\pm$1.22}} & \makecell{72.3\\\addtxt{$\pm$0.87}} & \makecell{48.6\\\addtxt{$\pm$1.02}} & \makecell{52.0\\\addtxt{$\pm$1.25}} & \makecell{48.6\\\addtxt{$\pm$1.10}} \\ \cline{2-8}
 & Subject 2 & \cellcolor{gray75}\makecell{90.0\\\addtxt{$\pm$0.15}} & \makecell{69.0\\\addtxt{$\pm$1.20}} & \makecell{86.6\\\addtxt{$\pm$0.95}} & \makecell{51.7\\\addtxt{$\pm$1.12}} & \makecell{48.6\\\addtxt{$\pm$1.30}} & \makecell{53.3\\\addtxt{$\pm$1.10}} \\ \cline{2-8}
 & Subject 3 & \cellcolor{gray75}\makecell{92.3\\\addtxt{$\pm$0.19}} & \makecell{65.3\\\addtxt{$\pm$1.15}} & \makecell{75.9\\\addtxt{$\pm$0.90}} & \makecell{49.2\\\addtxt{$\pm$1.08}} & \makecell{49.3\\\addtxt{$\pm$1.24}} & \makecell{49.1\\\addtxt{$\pm$1.22}} \\ \cline{2-8}
 & Subject 4 & \cellcolor{gray75}\makecell{93.7\\\addtxt{$\pm$0.17}} & \makecell{56.5\\\addtxt{$\pm$1.18}} & \makecell{68.4\\\addtxt{$\pm$1.00}} & \makecell{46.4\\\addtxt{$\pm$1.20}} & \makecell{47.5\\\addtxt{$\pm$1.14}} & \makecell{47.0\\\addtxt{$\pm$1.12}} \\ \cline{2-8}
 & Subject 5 & \cellcolor{gray75}\makecell{92.0\\\addtxt{$\pm$0.16}} & \makecell{72.5\\\addtxt{$\pm$1.20}} & \makecell{89.4\\\addtxt{$\pm$0.93}} & \makecell{50.4\\\addtxt{$\pm$1.08}} & \makecell{52.6\\\addtxt{$\pm$1.27}} & \makecell{54.9\\\addtxt{$\pm$1.20}} \\ \cline{2-8}
 & Subject 6 & \cellcolor{gray75}\makecell{94.2\\\addtxt{$\pm$0.20}} & \makecell{63.1\\\addtxt{$\pm$1.22}} & \makecell{76.0\\\addtxt{$\pm$0.88}} & \makecell{45.9\\\addtxt{$\pm$1.18}} & \makecell{46.4\\\addtxt{$\pm$1.10}} & \makecell{49.7\\\addtxt{$\pm$1.15}} \\ \cline{2-8}
 & Subject 7 & \cellcolor{gray75}\makecell{91.7\\\addtxt{$\pm$0.14}} & \makecell{63.6\\\addtxt{$\pm$1.15}} & \makecell{74.9\\\addtxt{$\pm$0.95}} & \makecell{46.6\\\addtxt{$\pm$1.12}} & \makecell{52.7\\\addtxt{$\pm$1.20}} & \makecell{50.1\\\addtxt{$\pm$1.15}} \\ \midrule

\multirow{7}{*}{Four emotions} & Subject 1 & \cellcolor{gray75}\makecell{78.5\\\addtxt{$\pm$0.15}} & \makecell{33.0\\\addtxt{$\pm$1.25}} & \makecell{58.0\\\addtxt{$\pm$1.10}} & \makecell{43.1\\\addtxt{$\pm$1.05}} & \makecell{42.0\\\addtxt{$\pm$1.18}} & \makecell{43.2\\\addtxt{$\pm$1.14}} \\ \cline{2-8}
 & Subject 2 & \cellcolor{gray75}\makecell{84.2\\\addtxt{$\pm$0.18}} & \makecell{55.5\\\addtxt{$\pm$1.10}} & \makecell{80.2\\\addtxt{$\pm$0.92}} & \makecell{56.5\\\addtxt{$\pm$0.88}} & \makecell{47.2\\\addtxt{$\pm$1.12}} & \makecell{45.5\\\addtxt{$\pm$1.08}} \\ \cline{2-8}
 & Subject 3 & \cellcolor{gray75}\makecell{82.9\\\addtxt{$\pm$0.15}} & \makecell{49.2\\\addtxt{$\pm$1.18}} & \makecell{45.1\\\addtxt{$\pm$1.24}} & \makecell{49.8\\\addtxt{$\pm$1.10}} & \makecell{45.2\\\addtxt{$\pm$1.20}} & \makecell{41.7\\\addtxt{$\pm$1.25}} \\ \cline{2-8}
 & Subject 4 & \cellcolor{gray75}\makecell{81.9\\\addtxt{$\pm$0.13}} & \makecell{34.6\\\addtxt{$\pm$1.20}} & \makecell{55.2\\\addtxt{$\pm$1.10}} & \makecell{45.9\\\addtxt{$\pm$1.18}} & \makecell{43.0\\\addtxt{$\pm$1.22}} & \makecell{45.9\\\addtxt{$\pm$1.17}} \\ \cline{2-8}
 & Subject 5 & \cellcolor{gray75}\makecell{86.7\\\addtxt{$\pm$0.12}} & \makecell{53.7\\\addtxt{$\pm$1.10}} & \makecell{77.5\\\addtxt{$\pm$0.85}} & \makecell{59.9\\\addtxt{$\pm$1.05}} & \makecell{49.6\\\addtxt{$\pm$1.20}} & \makecell{49.8\\\addtxt{$\pm$1.15}} \\ \cline{2-8}
 & Subject 6 & \cellcolor{gray75}\makecell{82.0\\\addtxt{$\pm$0.18}} & \makecell{50.8\\\addtxt{$\pm$1.15}} & \makecell{43.7\\\addtxt{$\pm$1.22}} & \makecell{53.4\\\addtxt{$\pm$1.08}} & \makecell{42.7\\\addtxt{$\pm$1.18}} & \makecell{38.7\\\addtxt{$\pm$1.22}} \\ \cline{2-8}
 & Subject 7 & \cellcolor{gray75}\makecell{83.1\\\addtxt{$\pm$0.16}} & \makecell{46.6\\\addtxt{$\pm$1.12}} & \makecell{49.6\\\addtxt{$\pm$1.14}} & \makecell{45.0\\\addtxt{$\pm$1.20}} & \makecell{49.1\\\addtxt{$\pm$1.15}} & \makecell{36.9\\\addtxt{$\pm$1.25}} \\

\bottomrule
\end{tabular}}
\end{table}

\subsection{UC3: Drowsiness detection}
\label{subsec:demonstrator-UC3}

The implementation performed in UC3 to detect drowsiness at the wheel consists \change{of}{in} using both EEG and Electrooculography (EOG) biosignals obtained from the BCI to detect when drivers are not suitable to drive, in addition to eye-tracking data to identify when drivers close their eyes. In particular, EOG uses electrodes near the eyes to measure the movement of the eyelid and thus help detect drowsy episodes. 

During the experiments, 21 subjects drove freely for around two hours, from whom 11 were men and 10 were women. Meanwhile, the framework acquired data from the BCI headset, obtaining both EEG and EOG signals. Focusing first on EEG, this UC used the Daisy board extension from OpenBCI, which augmented the number of available channels to 16, although reducing the sampling rate from 250 Hz to 125 Hz, enough for this UC. From them, 12 channels were used for EEG (O1, O2, P3, P4, Pz, T3, T4, T5, T6, C3, C5, Cz), while the remaining four were configured to act as EOG inputs. In particular, two EOG electrodes were placed at the exterior side of each eye, while the two remaining were placed vertically above the left eye. This setup ensured sufficient coverage to detect the movements of both eyelids. Moreover, the framework acquired eye-tracking data from the VR headset to identify when drivers closed their eyes. 

After the acquisition stage, the Data processing module of the framework cleaned EEG signals by employing a notch filter at 50 Hz, a bandpass filter between 1-30 Hz, downsampling to 60 Hz to reduce the complexity of subsequent stages, and, finally, ICA to remove artifacts. After having EEG signals with reduced noise, the framework divided the signals into eight-second epochs. From these data, the framework then extracted relevant features. For EEG, the framework calculated the PSD using the STFT method, obtaining information for each frequency band: delta, theta, alpha, beta, and gamma. Apart from PSD, the framework also calculated the percentage of eyelid closure (PERCLOS) obtained from eye-tracking data acquired in previous modules. PERCLOS is the division between the duration the eyes stay closed and the total time considered. Finally, the following features were transmitted to the AI-based detection: PSD of each EEG frequency band, raw EEG data, EOG data, and PERCLOS values. 

Finally, the framework trained regression and binary classification approaches to predict drowsiness based on the previous features. In particular, the models used the PERCLOS values as labels for predictions. The regression approach intended to predict the values of PERCLOS achieved, while the classification process determined whether a driver was drowsy. Specifically for the classification process, it was necessary to discretize the PERCLOS values to determine the thresholds that differentiate between awake and drowsy. This process followed the method defined by \cite{Hidalgo:drowsiness:2024}, based on a personalized discretization per subject, particularly using the moderate threshold defined in that research. For both classification approaches, the framework used the following ML algorithms: SVM, KNN, Decision Trees (DT), Gaussian Process (GP), and RF. Moreover, the experimentation isolated different features to determine which offered the most promising results. Specifically, the first configuration used raw EEG data, the second PSD values, and the third one, a combination of PSD and raw EOG data. \addtxt{It is also important to mention that PSD and EOG data were normalized using a MinMax scaler. Finally, the training process used tenfold cross-validation, using 70\% of the data for training and the remaining for testing, in addition to hyperparameter search to optimize the models.}

\begin{figure}[ht]
\begin{center}
\subfigure[Regression-based drowsiness identification.]{\includegraphics[width=\columnwidth]{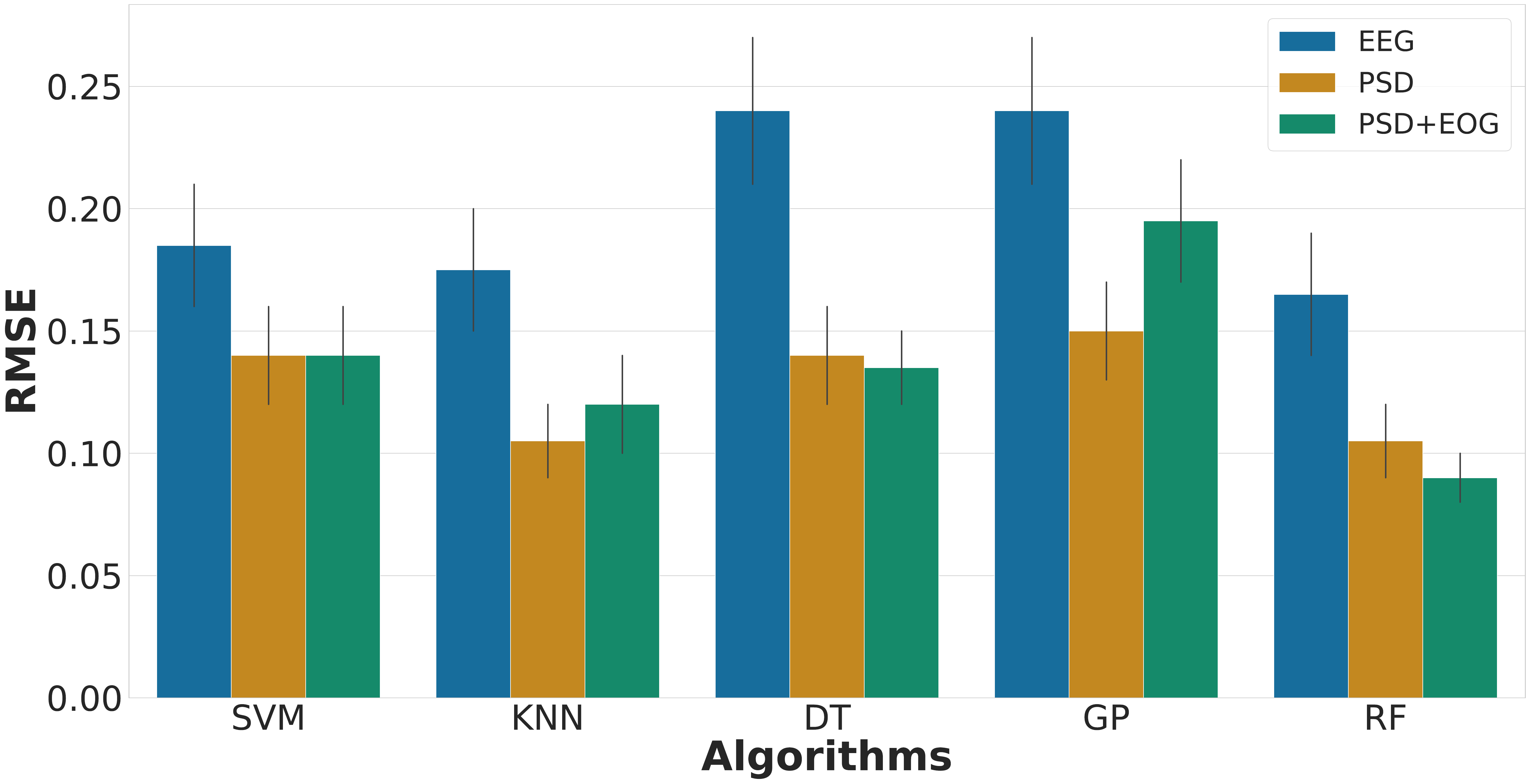}}
\subfigure[Drowsiness detection using data classification.]{\includegraphics[width=\columnwidth]{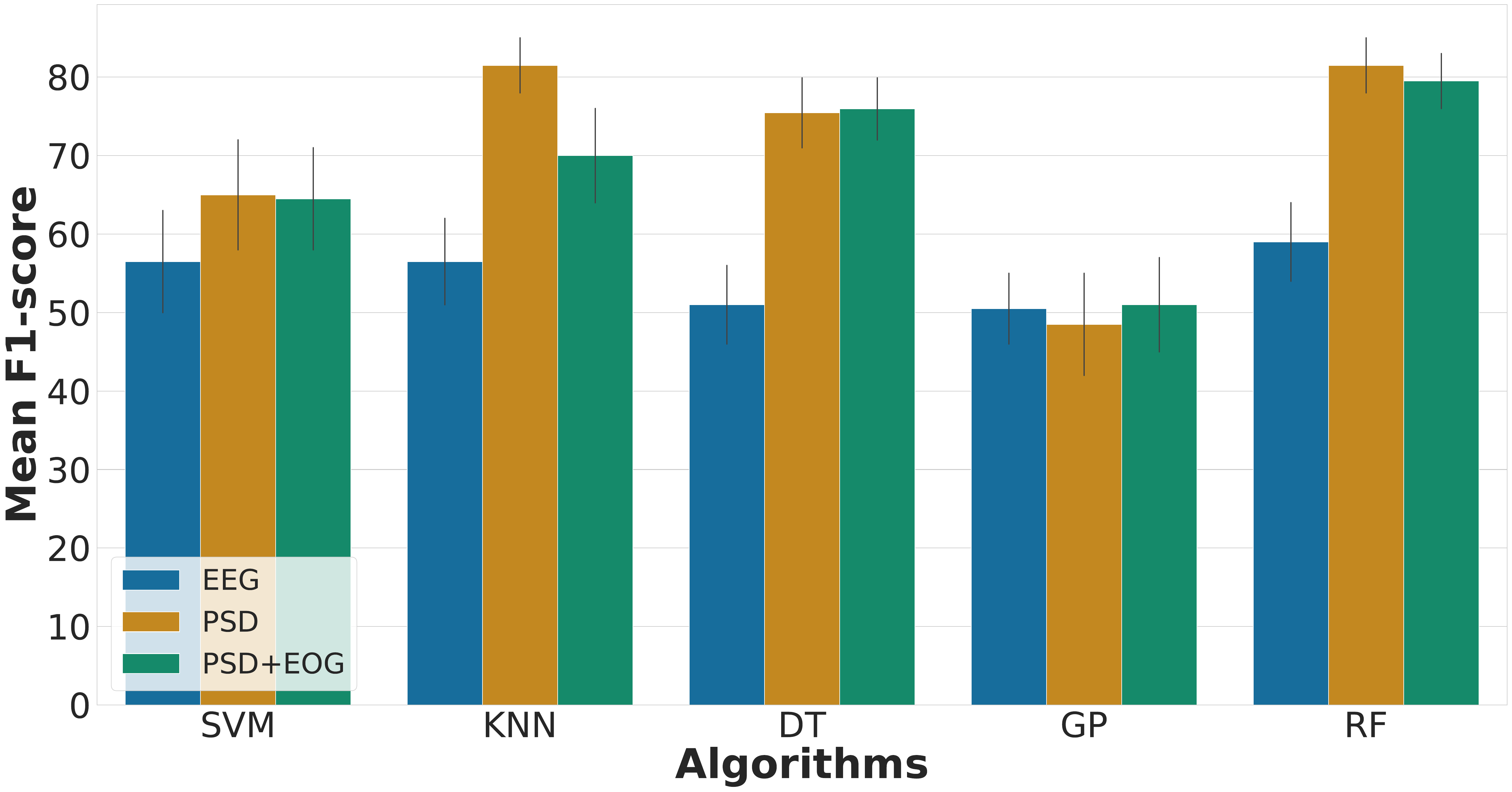}}
\end{center}
\caption{Results for UC3, presenting the performance in detecting drowsiness while simulated driving from both regression and classification approaches. Moreover, different features extracted from biosignals are studied.}
\label{fig:results-UC3}
\end{figure}

\figurename~\ref{fig:results-UC3} summarizes the results obtained in UC3 for each learning approach and algorithm tested. In both regression and classification approaches, RF offers the best results, although the performance of KNN is close. Focusing on the features studied, PSD offers the best performance for a regression approach, keeping the Root Mean Squared Error (RMSE) between 0.125 and 0.080 for all classifiers, closely followed by the joint use of PSD and EOG. Something similar happens using classification, where the most promising metric is PSD.

\subsection{UC4: Drivers' authentication}
\label{subsec:demonstrator-UC4}

The last UC aims to authenticate drivers using the simulation from two perspectives. The first approach aims to identify if the driver using the metaverse is indeed the intended and legitimate user (binary classification). In contrast, the second authentication approach focuses on detecting which user utilizes the metaverse within a set of possible drivers (multiclass classification), being useful for implementing access control mechanisms or billing for the services utilized.

Both \change{authentication}{biometric} alternatives present visual stimuli to the user to elicit a specific cerebral response following the oddball paradigm \cite{oddball}. This test visualizes images known to the subject within a set of unknown stimuli. Known images elicit a P300 potential, which can be measured as a voltage peak with a characteristic shape 
in occipital regions of the brain after 250-500 ms from the presentation of the stimuli \cite{p300wave}. It is essential to highlight that these P300 potentials present differences between subjects. Examples of research using P300 potentials for authentication are \cite{Rathi:auth_BCI:2021, LopezBernal:authentication:2023}. 

Before using the authentication mechanisms, it is necessary to obtain a sufficient amount of data to allow data classification. These data need to be as stable as possible; thus, the experiments were performed in a relaxed environment, without driving, using the BCI and the VR headsets. In particular, the data acquisition process for each subject was performed on two different days to provide variability to the EEG data gathered, using the Fp1, Fp2, C3, C4, P3, P4, O1, and O2 electrodes of the OpenBCI device on six men and four women while visualizing images. Ten tests were performed each day per subject, where a test consisted of presenting 40 times a known image within a set of 160 unknown and distinct images. In total, each test contained 200 images, and the stimuli used as known and unknown were different between tests. Furthermore, each image was displayed for one second. After training AI-based models, the users can use their previously trained models to authenticate. The framework presents known stimuli to the user before using the simulation. Once the authentication process has been completed, the user can start driving regularly. 

After acquiring EEG signals, the Data processing module applied a notch at 50 Hz, a bandpass filter between 1-17 Hz, and ICA to remove artifacts. Finally, this module extracts epochs from the EEG signals, selecting 0.1 seconds before a stimulus is displayed and 0.8 after its presentation. Then, the feature extraction process obtained statistical values from each epoch using a sliding window of 232 vectors (the total number of vectors per epoch), obtaining statistics such as mean, variance, median, or standard deviation. The considerations for this setup are inspired by previous work studying the best processing, feature extraction, and classification techniques for authenticating using P300 potentials \cite{LopezBernal:authentication:2023}.

Finally, the AI-based detection module uses several ML algorithms to predict binary and multiclass authentication, as presented in \tablename~\ref{tab:config-UC4}. As can be seen, each model uses specific classification techniques according to existing literature. \addtxt{Moreover, this UC used cross-validation to divide 80\% of the data for training purposes, whereas the 20\% remaining was used for testing. Finally, a hyperparameter search process was required to fine-tune the models.}

In this regard, \tablename~\ref{tab:results-UC4} presents the results obtained for both classification approaches. It is worth highlighting that RF offers a 100\% F1-score in both classification alternatives, highlighting the feasibility of introducing ML algorithms in metaverse scenarios to authenticate \addtxt{and identify} users using EEG signals.

\addtxt{The high performance achieved by RF can be explained by its ability to handle the complex, non-linear relationships inherent in EEG signals. Its ensemble approach, combining multiple decision trees, makes it robust to noise and data variability, which are common in EEG data collected across different sessions and subjects. The preprocessing performed ensures that only the most relevant and clean signals are retained, while the feature extraction process enhances the discriminative power of P300 potentials by providing useful features. In contrast, other algorithms, such as LDA and LR, are limited by their assumptions of linearity, while SVM often requires extensive parameter tuning to perform well.}

\begin{table}[!htb]
\centering
\caption{Configurations of the different models trained for UC4.}
\label{tab:config-UC4}
\resizebox{0.9\columnwidth}{!}{
\begin{tabular}{@{}lll@{}}
\toprule
\textbf{Algorithm} & \textbf{Classification techniques}\\ \midrule

Logistic Regression (LR) & Vectorizer, StandardScaler \\ \midrule
Linear Discriminant Analysis (LDA) & Vectorizer, XDawn \\ \midrule
Random Forest (RF) & Vectorizer \\ \midrule
Quadratic Discriminant Analysis (QDA) & Vectorizer \\ \midrule
Support Vector Machine (SVM) & Vectorizer \\ \midrule
k-Nearest Neighbors (KNN) & Vectorizer \\
\bottomrule
\end{tabular}}
\end{table}

\begin{table}[!htb]
\centering
\caption{Performance of the models trained in UC4 to authenticate drivers based on EEG data using the F1-score metric. Binary classification detects if a subject is legitimate when compared with the rest. In contrast, a multiclass approach determines to which subject an EEG signal sample corresponds.}
\label{tab:results-UC4}
\resizebox{\columnwidth}{!}{
\begin{tabular}{@{}cccccccc@{}}
\toprule
\textbf{Experiments} & \textbf{Subjects} & \textbf{LR} & \textbf{LDA} & \textbf{RF} & \textbf{QDA} & \textbf{SVM} & \textbf{KNN} \\ \midrule

\multirow{19}{*}{Binary} & Subject 1 & \makecell{80.0\\\addtxt{$\pm$0.00}} & \makecell{79.0\\\addtxt{$\pm$0.00}} & \cellcolor{gray75}\makecell{100.0\\\addtxt{$\pm$0.00}} & \makecell{71.4\\\addtxt{$\pm$1.67}} & \makecell{86.0\\\addtxt{$\pm$0.00}} & \makecell{88.0\\\addtxt{$\pm$0.00}} \\ \cline{2-8}
& Subject 2 & \makecell{80.0\\\addtxt{$\pm$0.00}} & \makecell{79.2\\\addtxt{$\pm$0.45}} & \cellcolor{gray75}\makecell{100.0\\\addtxt{$\pm$0.00}} & \makecell{69.2\\\addtxt{$\pm$2.05}} & \makecell{84.0\\\addtxt{$\pm$0.00}} & \makecell{86.0\\\addtxt{$\pm$0.00}} \\ \cline{2-8}
& Subject 3 & \makecell{78.0\\\addtxt{$\pm$0.00}} & \makecell{76.6\\\addtxt{$\pm$0.55}} & \cellcolor{gray75}\makecell{100.0\\\addtxt{$\pm$0.00}} & \makecell{71.2\\\addtxt{$\pm$0.45}} & \makecell{85.0\\\addtxt{$\pm$0.00}} & \makecell{85.0\\\addtxt{$\pm$0.00}} \\\cline{2-8}
& Subject 4 & \makecell{78.0\\\addtxt{$\pm$0.00}} & \makecell{76.0\\\addtxt{$\pm$0.00}} & \cellcolor{gray75}\makecell{100.0\\\addtxt{$\pm$0.00}} & \makecell{68.8\\\addtxt{$\pm$0.45}} & \makecell{83.0\\\addtxt{$\pm$0.00}} & \makecell{87.0\\\addtxt{$\pm$0.00}} \\ \cline{2-8}
& Subject 5 & \makecell{97.6\\\addtxt{$\pm$0.55}} & \makecell{97.0\\\addtxt{$\pm$0.00}} & \cellcolor{gray75}\makecell{100.0\\\addtxt{$\pm$0.00}} & \makecell{94.2\\\addtxt{$\pm$0.84}} & \makecell{99.0\\\addtxt{$\pm$0.00}} & \makecell{97.0\\\addtxt{$\pm$0.00}} \\\cline{2-8}
& Subject 6 & \makecell{73.0\\\addtxt{$\pm$0.00}} & \makecell{72.0\\\addtxt{$\pm$0.00}} & \cellcolor{gray75}\makecell{100.0\\\addtxt{$\pm$0.00}} & \makecell{61.0\\\addtxt{$\pm$1.41}} & \makecell{78.8\\\addtxt{$\pm$0.45}} & \makecell{86.0\\\addtxt{$\pm$0.00}} \\ \cline{2-8}
& Subject 7 & \makecell{69.0\\\addtxt{$\pm$0.00}} & \makecell{68.0\\\addtxt{$\pm$0.00}} & \cellcolor{gray75}\makecell{100.0\\\addtxt{$\pm$0.00}} & \makecell{65.4\\\addtxt{$\pm$0.55}} & \makecell{78.0\\\addtxt{$\pm$0.00}} & \makecell{84.0\\\addtxt{$\pm$0.00}} \\ \cline{2-8}
& Subject 8 & \makecell{75.0\\$\pm$0.00} & \makecell{74.0\\\addtxt{$\pm$0.00}} & \cellcolor{gray75}\makecell{100.0\\\addtxt{$\pm$0.00}} & \makecell{60.8\\\addtxt{$\pm$0.84}} & \makecell{82.8\\\addtxt{$\pm$0.45}} & \makecell{90.0\\\addtxt{$\pm$0.00}} \\ \cline{2-8}
& Subject 9 & \makecell{80.4\\\addtxt{$\pm$0.55}} & \makecell{79.8\\\addtxt{$\pm$0.45}} & \cellcolor{gray75}\makecell{100.0\\\addtxt{$\pm$0.00}} & \makecell{73.2\\\addtxt{$\pm$1.92}} & \makecell{85.0\\\addtxt{$\pm$0.00}} & \makecell{90.6\\\addtxt{$\pm$0.55}} \\\cline{2-8}
& Subject 10 & \makecell{71.0\\\addtxt{$\pm$0.00}} & \makecell{71.0\\\addtxt{$\pm$0.00}} & \cellcolor{gray75}\makecell{100.0\\\addtxt{$\pm$0.00}} & \makecell{61.8\\\addtxt{$\pm$1.10}} & \makecell{78.4\\\addtxt{$\pm$0.55}} & \makecell{87.0\\\addtxt{$\pm$0.00}} \\
\midrule

Multiclass & - & \makecell{49.0\\\addtxt{$\pm$0.00}} & \makecell{47.0\\\addtxt{$\pm$0.00}} & \cellcolor{gray75}\makecell{100.0\\\addtxt{$\pm$0.00}} & \makecell{42.0\\\addtxt{$\pm$0.89}} & \makecell{63.0\\\addtxt{$\pm$0.00}} & \makecell{87.0\\\addtxt{$\pm$0.00}} \\

\bottomrule

\end{tabular}}
\end{table}

\section{Results discussion}
\label{sec:discussion}

The results presented in the previous section indicate that including BCIs in metaverse scenarios is a promising research field, although it has not been widely explored. The four UCs defined obtained positive results concerning AI model performance, particularly for the demonstrator evaluated. However, further work is needed to elaborate robust commercial solutions. 

It is essential to note that BCI research is an active scientific field continuously improving, offering new technologies and techniques that improve existing commercial systems. In that way, the evolution of these interfaces will ease the integration of BCIs with the metaverse in the future. Moreover, the applicability of BCIs in particular knowledge domains, such as detecting emotions or drowsiness, is still open for improvement. 

The results obtained for UC1 to detect distractions at the wheel highlight that using BCIs in learning metaverses for driving can achieve more than 80\% F1-score for personalized models, sufficient for learning environments where small miss-classifications are not dramatic. Something similar happened when detecting the kind of distraction the driver suffered, achieving more than 70\% F1-score for the best configuration using RF as the classifier. 

In the same direction, identifying emotions performed in UC2 indicates that detecting two distinct emotions (no stimuli and angry) offers a performance of over \change{96}{93}\%. Moreover, considering three and four emotions presents excellent results (over 90\% and 78\%, respectively), even when performing complex and dynamic tasks, such as driving. Moreover, it is essential to note that some emotions, such as neutral or no stimuli, can be easily confused by classifiers (see \cite{Quiles:emotions:2023}). 

The use of BCIs to detect drowsiness is also a promising application area for the metaverse, specifically for learning scenarios, where professors can measure in real time if students are getting drowsy while interacting with new content. Particularly for driving metaverses, the results from UC3 obtained an F1-score close to 80\% for several algorithms. 

Finally, the results obtained from UC4 indicate that including authentication mechanisms based on EEG signals is a promising research area in the metaverse, capable of implementing access control mechanisms without the need for introducing passwords. In particular, the best performing algorithm obtained a 100\% F1-score for identifying both a legitimate user among a set of unauthorized drivers and detecting the specific user utilizing the metaverse without requiring providing profile details. 

\begin{figure*}[ht]
\begin{center}
\includegraphics[width=\textwidth]{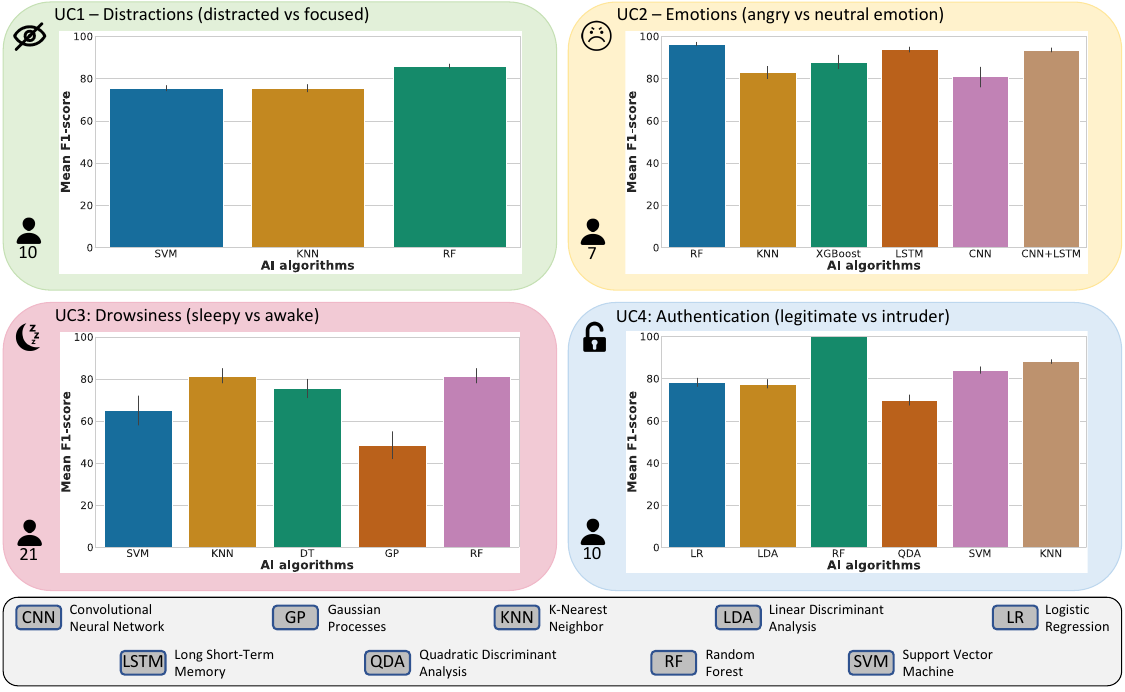}
\end{center}
\caption{Summary of the performance achieved by the proposed framework in four use cases of an educational driving metaverse scenario.}
\label{fig:results}
\end{figure*}

The results of this experimentation are summarized in \figurename~\ref{fig:results} per use case. Since the extensive experiments performed in each use case, this figure only highlights using binary classifiers to determine if the user is in good condition to drive. For UC1, the figure determines if the driver is distracted or focused. UC2 highlights whether the user is stressed or in a neutral mood, UC3 indicates if the driver is sleepy or awake, and UC4 specifies if the subject is legitimate or an intruder. 

As can be seen, all use cases achieved outstanding performances\addtxt{, being RF the algorithm presenting the best performance. As indicated in Section~\ref{subsec:demonstrator-UC4} for UC4, this difference can be explained by the nature of EEG data and how RF, among ML algorithms, is capable of identifying patterns that other algorithms, due to their linear or complex behavior, cannot identify with the amount of data provided or the configurations tested. Moreover, although DL algorithms performed well in UC2, they did not outperform RF in this paper. The limited number of subjects can explain this situation and, thus, the constraints in the data generated in the experimentation. Increasing the data used to train the models could considerably improve the performance of DL algorithms.}

\change{These}{In summary, the previous} results demonstrate the feasibility of using BCIs in educational driving environments to assess users' optimal driving conditions and authenticate them in the metaverse\addtxt{, although additional research in this direction is still needed}.

\section{Trends in BCIs and potential benefits for their inclusion in metaverse scenarios}
\label{sec:trends_challenges}
To understand how BCIs can contribute to the metaverse and the actions to be performed for this achievement, it is relevant to analyze the current trends in BCIs from both research and commercial perspectives.

\textbf{BCIs are undergoing a technological revolution}. These interfaces are moving to a wireless scenario since they provide better versatility during their use, an essential aspect of their inclusion in the metaverse to operate in diverse environments. Moreover, the technological advances of the last decade have generated an increase in BCI capabilities and a cost reduction \cite{Lopez_Bernal:cyberattacks_implants:2020}. These aspects induce the development of technologies that are more attractive and accessible to final users in the metaverse.

\textbf{Non-invasive BCIs are gaining popularity}. Non-invasive BCIs have gained popularity due to a reduced neurostimulation risk and a complete suppression of neural data acquisition harm \cite{Lopez_Bernal:cyberBCI:2021}. Nevertheless, this comes with a trade-off in terms of resolution. In neurostimulation, these devices are not as precise as invasive ones, covering broader brain regions and not being possible to target individual neurons. In data acquisition, these BCIs can only receive aggregated data from broad neuronal populations, thus defining an upper threshold for the progress of these BCI technologies. Although non-invasive BCIs are more suitable nowadays for the metaverse, the current technological trend could provide invasive systems with minimum risks and enormous benefits for metaverse users in the future.

\textbf{Broad commercial activity between BCIs and the metaverse}. Big technological corporations, such as Meta or Snap, have recently acquired BCI companies to integrate these interfaces into their hardware developments focused on metaverse solutions. Valve, a video game company producing both software and hardware, is closely collaborating with OpenBCI, a BCI company producing low-cost EEG devices aiming to democratize the access to these devices. Finally, companies such as Neuralink aim to develop next-generation BCIs to integrate our minds with the virtual world.

\textbf{Futuristic BCI projects for the metaverse tend towards invasive systems}. The invasiveness of BCIs is one of the key aspects of these interfaces, closely related to neural data quality. Invasive BCIs for either data acquisition or neurostimulation have considerably progressed in medical scenarios, achieving better miniaturization of technology and electrodes. This advance aims for a better study and control of the brain by increasing both temporal and spatial resolutions \cite{Ramadan:controlSignalsReview:2017}. Thus, this evolution is crucial to achieving a broad coverage of the brain, where the ultimate goal is to target tiny neuronal populations, or even individual neurons, as in the case of projects such as Neuralink or Synchron. In fact, the scientific literature highlights the necessity of novel BCI approaches to achieve completely immersive metaverses \cite{Dwivedi:metaverse_multidisciplinary:2022}. 

\textbf{Novel artificial intelligence techniques are key in BCI}. In recent years, BCI research has considerably improved the reliability and efficacy of existing solutions by using intelligent approaches, such as machine learning and deep learning, also presenting a trend in creating general models that different users can use \cite{Zhang:BCI_AI:2020}. Literature is also active in preserving users' data privacy when training collaborative models, where federated learning is promising \cite{Ju:BCI_FL:2020, MartinezBeltran:DFL_survey:2023}.

Moreover, the trend of these interfaces anticipates a series of challenges that BCI research needs to resolve for their complete integration within the metaverse.

\textbf{Limited understanding of the brain}. Although promising advances in neuroscience are made every day, they are still insufficient to achieve such ambitious objectives. For that, multidisciplinary research teams are essential to explore the brain better, supported by cutting-edge computational developments.

\textbf{Reduced data quality}. More work is required to reduce the noise captured by electrodes, which can result in errors when decoding the meaning of these signals \cite{Lebedev:BrainMachineIF:2017}. Additionally, research will require improving both temporal and spatial resolutions, as they are currently insufficient for futuristic metaverse applications. The improvement and development of technologies with better resolutions can solve these issues. 

\textbf{Limited capabilities in neurostimulation}. Both invasive and non-invasive neurostimulation systems face a considerable challenge in reducing their risks of neural tissue damage as they will be one of the keys to their implantation in the metaverse \cite{Khabarova:dbsParkinson:2018}. Moreover, current implantable BCI systems for stimulation are immature for their use in the metaverse, thus demanding an effort in both research and innovation dimensions. Thus, novel technologies with nanoscale electrodes will be key for the metaverse. 

\textbf{Lack of interoperability}. Current BCI deployments are incompatible since companies focus on creating systems for solving particular problems, not being compatible with other products in most cases. This situation is also aggravated by a lack of BCI standards, where data formats and communication links are not homogeneous in the proposed solutions. This challenge can be eased by a standardization effort to unify BCI technologies and solutions \cite{Lopez_Bernal:cyberBCI:2021}.

\textbf{Impossibility to extend functionality}. BCIs are not commonly extensible as they cannot incorporate new functionality. This feature is essential for their use in changing environments, such as the metaverse. This aspect is even more critical when considering the cybersecurity prism since these devices cannot incorporate new protecting mechanisms when previous techniques become obsolete. Thus, BCIs must follow a modular architecture, also considering security-by-default strategies \cite{Lopez_Bernal:cyberBCI:2021}.

\section{Concerns of the integration of BCIs into the metaverse}
\label{sec:concerns}
Despite the advantages BCIs can introduce to the metaverse, their application raises several concerns that must be carefully addressed. These considerations are grouped into the following categories: accessibility, user integration, privacy, cybersecurity, users' safety, and ethics. 

\textbf{BCIs as an accessibility issue}. Although these interfaces could introduce revolutionary capabilities to our daily lives when meeting the metaverse, they could also cause difficulties in the access to information. The evolution in modern technological systems, such as personal computers and smartphones, already generated a tremendous impact in society, although it took place during a relatively large period. However, the rapid changes of technology nowadays, particularly in metaverse scenarios, could generate a dramatic impact on society. 

\addtxt{\textbf{Side effects}. The use of existing implantable BCIs can generate undesired effects, especially during surgery, such as bleeding or tissue damage. However, the evolution of implantable BCIs toward the miniaturization of electrodes aims to reduce these drawbacks \cite{Leuthardt:RisksBCI:2021}. In contrast, the combination of non-invasive BCI systems with VR generates different adverse effects, such as dizziness, vertigo, headache, or nausea, as documented by literature addressing the drawbacks of VR systems \cite{Biswas:cybersickness:2024}.} 

\textbf{Inclusion of BCI users in the metaverse}. The analysis regarding BCI users in the metaverse can be considered from two dimensions. First, employing BCIs might ease the inclusion of users with disabilities, such as patients with motor, vision or hearing impairing. These users could find novel ways to socialize and communicate, or even surpass the restrictions caused by their health conditions. Despite these benefits, the technological revolution introduced by novel-generation BCIs could increase social inequity based on diverse factors, such as age or purchasing power \cite{Liu2024}. Because of that, the implantation roadmap of these novel systems should be carefully studied.

\textbf{Privacy - Continuous data acquisition}. The metaverse is characterized by continuous and pervasive user data acquisition. This aspect is critical since vast amounts of data are stored and transmitted to the Internet that could be used for malicious purposes \cite{Wang:survey_security:2022, Wang:survey:2023}. For example, unauthorized attackers could access these data to sell them to third parties, blackmail users, or perform social profiling \cite{Wang:survey_security:2022, DiPietro:security_conf:2021}. The situation worsens when using BCIs if illegitimate parties obtain users’ neural data and neurostimulation patterns due to the criticality of these data.

\textbf{Privacy - Lack of users' privacy}. Previous work has identified confidentiality issues related to BCIs affecting users’ privacy \cite{Liu2024}. Applications using P300 potentials to discriminate between known and unknown visual stimuli, widely used in the BCI domain for different purposes like mental spellers, can be used by attackers to obtain sensitive users’ data, such as faces, locations, or PINs, as well as their political or sexual orientation \cite{QuilesPerez:privacy_BCI:2021}. This circumstance is critical since attackers could use the acquired information to spy and blackmail metaverse users.

\textbf{Cybersecurity - User impersonation and data tampering}. The literature has specified different categories of threats that affect the metaverse \cite{Wang:survey_security:2022, DiPietro:security_conf:2021}. First, attackers could impersonate legitimate users in simulations using BCIs in charge of authentication by replicating previously recorded brain signals. Moreover, data tampering attacks, aiming to modify data maliciously, can be performed over brain signals to affect the integrity and availability of the system \cite{MartinezBeltran:BCI_noise:2022}.

\textbf{Cybersecurity - Application of traditional cyberattacks}. Apart from the previous threats, a wide variety of cyberattacks can be applied to BCIs in each of their functioning stages \cite{Lopez_Bernal:cyberBCI:2021}. In a nutshell, most of the cyberattacks applicable to computers are also applicable to BCIs, generating a wide range of adverse impacts on data and service integrity, confidentiality, and availability. 

\textbf{Cybersecurity - Neural cyberattacks}. Additionally, novel neurostimulation BCIs, characterized by the miniaturization of electrodes and increased brain coverage, open the door for attacks altering spontaneous neuronal activity, known as neural cyberattacks \cite{Lopez_Bernal:cyberattacks_implants:2020, Lopez_Bernal:jamming:2022, LopezBernal:taxonomy_attacks:2021}. This situation could allow a modification of the stimuli sent by metaverse users, altering their experience within the simulation.

\textbf{Users' safety - Attacks against the brain}. Apart from affecting the simulation, attacks can aim to damage users’ physical integrity directly \cite{Li:bciApplications:2015}. While using BCIs, particularly those used for neurostimulation, attackers could aim to overstimulate targeted brain regions or inhibit them by disrupting the regular activity of the brain. These effects can be aggravated in metaverse scenarios if a broad brain coverage is achieved, as neural cyberattacks can affect the brain with single-neuron resolution. The damage caused by these threats could even recreate the effects of neurodegenerative diseases, although more research is required in this direction \cite{Lopez_Bernal:jamming:2022}. 

\textbf{Ethics - Misuse of BCIs}. On the one hand, neural data with sufficient resolution could provide extremely sensitive information about the subject, even outside its application scope \cite{Park:metaverse_challenges:2022}. On the other hand, the potential damage that can be caused to subjects’ safety could be employed by users in terrorist activities to damage, or even kill, people with opposite ideologies or rival countries. Thus, using BCIs within the metaverse for the long term generates risks that must be considered before their final implementation.

\section{Conclusions}
\label{sec:conclusions}
BCIs emerge as promising solutions in the metaverse, enabling the control of elements with the mind and providing better immersion and more realistic sensations. Since the inclusion of BCIs in the metaverse is not explored in depth, this work presents the first analysis of the current applicability of BCIs to the metaverse, also studying its medium and long terms. In particular, this paper studies their utilization in popular metaverse application scenarios, such as video games or education environments. It also reviews their use for incorporating the human senses into the virtual world. 

Furthermore, the present manuscript describes the design and implementation of a general framework capable of interconnecting BCIs with other sensors and actuators to provide more immersive simulated experiences. This framework can be extended to incorporate new functionality based on its modular design, such as adding more data processing techniques or AI algorithms. To demonstrate the integration of the framework with non-invasive BCIs, this paper has validated the platform on four well-known use cases from the BCI domain in a driving metaverse for learning: detection of distractions, identification of emotions, drowsiness assessment, and user authentication. To provide a more immersive experience, a VR headset and a steering wheel with pedals have also been added. This demonstrator highlights that BCIs are promising technologies for their inclusion in the early stages of the metaverse, although further research is needed to achieve more immersive scenarios.

This work also introduces the current trend of BCIs and how this evolution can contribute to the metaverse. Although these advances predict a promising future, they also present several challenges that must be addressed for their complete implementation in the metaverse. Finally, several concerns from different perspectives are presented, highlighting that despite the advantages of BCI technologies, several aspects are critical to addressing before their full implantation. 

Future work could comprehensively analyze the feasibility of including prospecting BCI technologies in metaverse scenarios, considering their current capabilities and limitations. Additionally, future work could focus on diverse aspects of including BCIs not explored nowadays, such as user experience or the customization of the systems based on subjects' patterns. Moreover, the harmonization of standards and regulations is an open challenge, specifically for safety, privacy, and ethical issues. \change{Finally}{Furthermore}, there is a lack of detailed works studying cybersecurity aspects applicable to BCIs on the metaverse. \addtxt{Finally, there is an opportunity to assess the applicability of DL algorithms in all UCs defined, where this work only used these techniques in UC2.}

\section*{Acknowledgments}

\section*{Funding}
This work has been partially supported by (a) 21628/FPI/21 and 21629/FPI/21 Fundaci\'on S\'eneca, cofunded by Bit \& Brain Technologies S.L. Regi\'on de Murcia (Spain), (b) the strategic project CDL-TALENTUM from the Spanish National Institute of Cybersecurity (INCIBE) and by the Recovery, Transformation and Resilience Plan, Next Generation EU, (c) the Swiss Federal Office for Defense Procurement (armasuisse) with the CyberTracer (CYD-C-2020003) project, and (d) the University of Zürich UZH.

\bibliography{references}

\newcommand{\noop}[1]{}
\begin{thebibliography}{10}
\expandafter\ifx\csname url\endcsname\relax
  \def\url#1{\texttt{#1}}\fi
\expandafter\ifx\csname urlprefix\endcsname\relax\def\urlprefix{URL }\fi
\expandafter\ifx\csname href\endcsname\relax
  \def\href#1#2{#2} \def\path#1{#1}\fi

\bibitem{Zhang:education:2022}
X.~Zhang, Y.~Chen, L.~Hu, Y.~Wang, The metaverse in education: Definition, framework, features, potential applications, challenges, and future research topics, Frontiers in Psychology 13 (2022).
\newblock \href {https://doi.org/10.3389/fpsyg.2022.1016300} {\path{doi:10.3389/fpsyg.2022.1016300}}.

\bibitem{Mystakidis:metaverse:2022}
S.~Mystakidis, Metaverse, Encyclopedia 2~(1) (2022) 486--497.
\newblock \href {https://doi.org/10.3390/encyclopedia2010031} {\path{doi:10.3390/encyclopedia2010031}}.

\bibitem{Wang:survey_security:2022}
Y.~Wang, Z.~Su, N.~Zhang, R.~Xing, D.~Liu, T.~H. Luan, X.~Shen, A survey on metaverse: Fundamentals, security, and privacy, IEEE Communications Surveys \& Tutorials (2022) 1--1\href {https://doi.org/10.1109/COMST.2022.3202047} {\path{doi:10.1109/COMST.2022.3202047}}.

\bibitem{Narin:analysis_metaverse:2021}
N.~Gökçe~Narin, A content analysis of the metaverse articles, Journal of Metaverse 1~(1) (2021) 17 -- 24.

\bibitem{Njoku:metaverse_driving:2022}
J.~N. Njoku, C.~I. Nwakanma, G.~C. Amaizu, D.-S. Kim, Prospects and challenges of metaverse application in data-driven intelligent transportation systems, IET Intelligent Transport Systems (2022) 21\href {https://doi.org/10.1049/itr2.12252} {\path{doi:10.1049/itr2.12252}}.

\bibitem{Park:metaverse_challenges:2022}
S.-M. Park, Y.-G. Kim, A metaverse: Taxonomy, components, applications, and open challenges, IEEE Access 10 (2022) 4209--4251.
\newblock \href {https://doi.org/10.1109/ACCESS.2021.3140175} {\path{doi:10.1109/ACCESS.2021.3140175}}.

\bibitem{DiPietro:security_conf:2021}
R.~Di~Pietro, S.~Cresci, Metaverse: Security and privacy issues, in: 2021 Third IEEE International Conference on Trust, Privacy and Security in Intelligent Systems and Applications (TPS-ISA), IEEE, Atlanta, GA, USA, 2021, pp. 281--288.
\newblock \href {https://doi.org/10.1109/TPSISA52974.2021.00032} {\path{doi:10.1109/TPSISA52974.2021.00032}}.

\bibitem{Wang:survey:2023}
H.~Wang, H.~Ning, Y.~Lin, W.~Wang, S.~Dhelim, F.~Farha, J.~Ding, M.~Daneshmand, A survey on the metaverse: The state-of-the-art, technologies, applications, and challenges, IEEE Internet of Things Journal 10~(16) (2023) 14671--14688.
\newblock \href {https://doi.org/10.1109/JIOT.2023.3278329} {\path{doi:10.1109/JIOT.2023.3278329}}.

\bibitem{Khabarova:dbsParkinson:2018}
E.~Khabarova, N.~Denisova, A.~Dmitriev, K.~Slavin, L.~Verhagen~Metman, Deep brain stimulation of the subthalamic nucleus in patients with parkinson disease with prior pallidotomy or thalamotomy, Brain Sciences 8~(4) (2018) 66.

\bibitem{Li:bciApplications:2015}
Q.~Li, D.~Ding, M.~Conti, {Brain-Computer Interface applications: Security and privacy challenges}, in: 2015 IEEE Conference on Communications and Network Security (CNS), IEEE, San Francisco, CA, USA, 2015, pp. 663--666.

\bibitem{Papanastasiou:BCI_improvement:2022}
G.~Papanastasiou, A.~Drigas, C.~Skianis, M.~Lytras, Brain computer interface based applications for training and rehabilitation of students with neurodevelopmental disorders. a literature review, Heliyon 6~(9) (2020) e04250.
\newblock \href {https://doi.org/10.1016/j.heliyon.2020.e04250} {\path{doi:10.1016/j.heliyon.2020.e04250}}.

\bibitem{Jiang:btb_brainet:2019}
L.~Jiang, A.~Stocco, D.~M. Losey, J.~A. Abernethy, C.~S. Prat, R.~P.~N. Rao, Brainnet: A multi-person brain-to-brain interface for direct collaboration between brains, Scientific Reports 9~(1) (2019) 6115.
\newblock \href {https://doi.org/10.1038/s41598-019-41895-7} {\path{doi:10.1038/s41598-019-41895-7}}.

\bibitem{Edward:SoftSkills:2022}
S.~Edward, K.~J. Hyun, The metaverse and video games: Merging media to improve soft skills training, Journal of Internet Computing and Services 23~(1) (2022) 69--76.

\bibitem{Haihan:metaverse_campus:2021}
H.~Duan, J.~Li, S.~Fan, Z.~Lin, X.~Wu, W.~Cai, Metaverse for social good: A university campus prototype, in: Proceedings of the 29th ACM International Conference on Multimedia, MM '21, Association for Computing Machinery, New York, NY, USA, 2021, p. 153–161.
\newblock \href {https://doi.org/10.1145/3474085.3479238} {\path{doi:10.1145/3474085.3479238}}.

\bibitem{Contreras:education:2022}
G.~S. Contreras, A.~H. González, M.~I.~S. Fernández, C.~B.~M. Cepa, J.~C.~Z. Escobar, The importance of the application of the metaverse in education, Modern Applied Science 16 (2022) 34.
\newblock \href {https://doi.org/10.5539/MAS.V16N3P34} {\path{doi:10.5539/MAS.V16N3P34}}.

\bibitem{Barrera:marketing:2023}
K.~{Giang Barrera}, D.~Shah, Marketing in the metaverse: Conceptual understanding, framework, and research agenda, Journal of Business Research 155 (2023) 113420.
\newblock \href {https://doi.org/https://doi.org/10.1016/j.jbusres.2022.113420} {\path{doi:https://doi.org/10.1016/j.jbusres.2022.113420}}.

\bibitem{Buhalis:tourism:2023}
D.~Buhalis, D.~Leung, M.~Lin, Metaverse as a disruptive technology revolutionising tourism management and marketing, Tourism Management 97 (2023) 104724.
\newblock \href {https://doi.org/https://doi.org/10.1016/j.tourman.2023.104724} {\path{doi:https://doi.org/10.1016/j.tourman.2023.104724}}.

\bibitem{Yang:medicine:2022}
D.~Yang, J.~Zhou, R.~Chen, Y.~Song, Z.~Song, X.~Zhang, Q.~Wang, K.~Wang, C.~Zhou, J.~Sun, L.~Zhang, L.~Bai, Y.~Wang, X.~Wang, Y.~Lu, H.~Xin, C.~A. Powell, C.~Thüemmler, N.~H. Chavannes, W.~Chen, L.~Wu, C.~Bai, Expert consensus on the metaverse in medicine, Clinical eHealth 5 (2022) 1--9.
\newblock \href {https://doi.org/https://doi.org/10.1016/j.ceh.2022.02.001} {\path{doi:https://doi.org/10.1016/j.ceh.2022.02.001}}.

\bibitem{Wu:emergency_medical:2022}
T.-C. Wu, C.-T.~B. Ho, A scoping review of metaverse in emergency medicine, Australasian Emergency Care (2022).
\newblock \href {https://doi.org/https://doi.org/10.1016/j.auec.2022.08.002} {\path{doi:https://doi.org/10.1016/j.auec.2022.08.002}}.

\bibitem{Zhou:epilepsy_detection:2018}
M.~Zhou, C.~Tian, R.~Cao, B.~Wang, Y.~Niu, T.~Hu, H.~Guo, J.~Xiang, Epileptic seizure detection based on eeg signals and cnn, Frontiers in Neuroinformatics 12 (2018).
\newblock \href {https://doi.org/10.3389/fninf.2018.00095} {\path{doi:10.3389/fninf.2018.00095}}.

\bibitem{Lebedev:BrainMachineIF:2017}
M.~A. Lebedev, M.~A.~L. Nicolelis, {Brain-Machine Interfaces: From Basic Science to Neuroprostheses and Neurorehabilitation}, Physiological Reviews 97~(2) (2017) 767--837.

\bibitem{Xu:poc_authentication:2021}
T.~Xu, H.~Wang, G.~Lu, F.~Wan, M.~Deng, P.~Qi, A.~Bezerianos, C.~Guan, Y.~Sun, E-key: an eeg-based biometric authentication and driving fatigue detection system, IEEE Transactions on Affective Computing (2021) 1--1\href {https://doi.org/10.1109/TAFFC.2021.3133443} {\path{doi:10.1109/TAFFC.2021.3133443}}.

\bibitem{DeOliveiraJunior:BtI:2018}
W.~G. de~Oliveira~J{\'u}nior, J.~M. de~Oliveira, R.~Munoz, V.~H.~C. de~Albuquerque, A proposal for internet of smart home things based on bci system to aid patients with amyotrophic lateral sclerosis, Neural Computing and Applications 32~(15) (2020) 11007--11017.

\bibitem{Musk:neuralink:2019}
E.~Musk, An integrated brain-machine interface platform with thousands of channels, J Med Internet Res 21~(10) (2019) e16194.
\newblock \href {https://doi.org/10.2196/16194} {\path{doi:10.2196/16194}}.

\bibitem{Opie:Synchron:2018}
N.~L. Opie, S.~E. John, G.~S. Rind, S.~M. Ronayne, Y.~T. Wong, G.~Gerboni, P.~E. Yoo, T.~J.~H. Lovell, T.~C.~M. Scordas, S.~L. Wilson, A.~Dornom, T.~Vale, T.~J. O'Brien, D.~B. Grayden, C.~N. May, T.~J. Oxley, Focal stimulation of the sheep motor cortex with a chronically implanted minimally invasive electrode array mounted on an endovascular stent, Nature Biomedical Engineering 2~(12) (2018) 907--914.
\newblock \href {https://doi.org/10.1038/s41551-018-0321-z} {\path{doi:10.1038/s41551-018-0321-z}}.

\bibitem{Kerous:games_BCI:2018}
B.~Kerous, F.~Skola, F.~Liarokapis, Eeg-based bci and video games: a progress report, Virtual Reality 22~(2) (2018) 119--135.
\newblock \href {https://doi.org/10.1007/s10055-017-0328-x} {\path{doi:10.1007/s10055-017-0328-x}}.

\bibitem{QuilesPerez:privacy_BCI:2021}
M.~Quiles~P{\'e}rez, E.~T. Mart{\'i}nez~Beltr{\'a}n, S.~L{\'o}pez~Bernal, A.~Huertas~Celdr{\'a}n, G.~Mart{\'i}nez~P{\'e}rez, Breaching subjects' thoughts privacy: A study with visual stimuli and brain-computer interfaces, Journal of Healthcare Engineering 2021 (2021) 5517637.
\newblock \href {https://doi.org/10.1155/2021/5517637} {\path{doi:10.1155/2021/5517637}}.

\bibitem{Halim:poc_emotions:2020}
Z.~Halim, M.~Rehan, On identification of driving-induced stress using electroencephalogram signals: A framework based on wearable safety-critical scheme and machine learning, Information Fusion 53 (2020) 66--79.

\bibitem{Quiles:emotions:2023}
M.~Quiles~P{\'e}rez, E.~T. Mart{\'i}nez~Beltr{\'a}n, S.~L{\'o}pez~Bernal, G.~Mart{\'i}nez~P{\'e}rez, A.~Huertas~Celdr{\'a}n, Analyzing the impact of driving tasks when detecting emotions through brain--computer interfaces, Neural Computing and Applications 35~(12) (2023) 8883--8901.
\newblock \href {https://doi.org/10.1007/s00521-023-08343-0} {\path{doi:10.1007/s00521-023-08343-0}}.

\bibitem{Quiles:neuromarketing:2023}
M.~Q. Pérez, E.~T.~M. Beltrán, S.~L. Bernal, E.~H. Prat, L.~M.~D. Campo, L.~F. Maimó, A.~H. Celdrán, Data fusion in neuromarketing: Multimodal analysis of biosignals, lifecycle stages, current advances, datasets, trends, and challenges (2023).
\newblock \href {http://arxiv.org/abs/2209.00993} {\path{arXiv:2209.00993}}.

\bibitem{Dieter:hearing_optogenetics:2020}
A.~Dieter, D.~Keppeler, T.~Moser, Towards the optical cochlear implant: optogenetic approaches for hearing restoration, EMBO Molecular Medicine 12~(4) (2020) e11618.
\newblock \href {https://doi.org/10.15252/emmm.201911618} {\path{doi:10.15252/emmm.201911618}}.

\bibitem{Eggermont:hearing_midbrain:2017}
J.~J. Eggermont, Chapter 12 - auditory brainstem and midbrain implants, in: J.~J. Eggermont (Ed.), Hearing Loss, Academic Press, 2017, pp. 351--366.
\newblock \href {https://doi.org/10.1016/B978-0-12-805398-0.00012-8} {\path{doi:10.1016/B978-0-12-805398-0.00012-8}}.

\bibitem{Ganzer:touch_BCI:2020}
P.~D. Ganzer, S.~C. Colachis, M.~A. Schwemmer, D.~A. Friedenberg, C.~F. Dunlap, C.~E. Swiftney, A.~F. Jacobowitz, D.~J. Weber, M.~A. Bockbrader, G.~Sharma, Restoring the sense of touch using a sensorimotor demultiplexing neural interface, Cell 181 (2020) 763--773.e12.
\newblock \href {https://doi.org/10.1016/J.CELL.2020.03.054/ATTACHMENT/A36E1150-3D76-4E3F-96B5-21AC511A72C9/MMC2.MP4} {\path{doi:10.1016/J.CELL.2020.03.054/ATTACHMENT/A36E1150-3D76-4E3F-96B5-21AC511A72C9/MMC2.MP4}}.

\bibitem{Flesher:touch_BCI:2021}
S.~N. Flesher, J.~E. Downey, J.~M. Weiss, C.~L. Hughes, A.~J. Herrera, E.~C. Tyler-Kabara, M.~L. Boninger, J.~L. Collinger, R.~A. Gaunt, A brain-computer interface that evokes tactile sensations improves robotic arm control, Science 372~(6544) (2021) 831--836.
\newblock \href {https://doi.org/10.1126/science.abd0380} {\path{doi:10.1126/science.abd0380}}.

\bibitem{Niketeghad:sight:2019}
S.~Niketeghad, N.~Pouratian, Brain machine interfaces for vision restoration: The current state of cortical visual prosthetics, Neurotherapeutics 16~(1) (2019) 134--143.
\newblock \href {https://doi.org/10.1007/s13311-018-0660-1} {\path{doi:10.1007/s13311-018-0660-1}}.

\bibitem{Miyashita:taste_TV:2021}
H.~Miyashita, Tttv (taste the tv): Taste presentation display for “licking the screen” using a rolling transparent sheet and a mixture of liquid sprays, in: The Adjunct Publication of the 34th Annual ACM Symposium on User Interface Software and Technology, UIST '21, Association for Computing Machinery, New York, NY, USA, 2021, p. 37–40.
\newblock \href {https://doi.org/10.1145/3474349.3480223} {\path{doi:10.1145/3474349.3480223}}.

\bibitem{Anbarasan:taste_BCI:2022}
R.~Anbarasan, D.~Gomez~Carmona, R.~Mahendran, Human taste-perception: Brain computer interface (bci) and its application as an engineering tool for taste-driven sensory studies, Food Engineering Reviews (2022) 1--27\href {https://doi.org/10.1007/s12393-022-09308-0} {\path{doi:10.1007/s12393-022-09308-0}}.

\bibitem{Guimaraes:olfactory_vr:2022}
M.~de~Paiva~Guimar{\~a}es, J.~M. Martins, D.~R.~C. Dias, R.~d. F.~R. Guimar{\~a}es, B.~B. Gnecco, An olfactory display for virtual reality glasses, Multimedia Systems (2022) 1--11\href {https://doi.org/10.1007/s00530-022-00908-8} {\path{doi:10.1007/s00530-022-00908-8}}.

\bibitem{Yang:olfactory_BCI:2022}
Q.~Yang, G.~Zhou, T.~Noto, J.~W. Templer, S.~U. Schuele, J.~M. Rosenow, G.~Lane, C.~Zelano, Smell-induced gamma oscillations in human olfactory cortex are required for accurate perception of odor identity, PLOS Biology 20 (2022) e3001509.
\newblock \href {https://doi.org/10.1371/JOURNAL.PBIO.3001509} {\path{doi:10.1371/JOURNAL.PBIO.3001509}}.

\bibitem{Holbrook:smell_BCI:2020}
E.~H. Holbrook, D.~H. Coelho, Cranial nerve stimulation for olfaction (cranial nerve 1), Otolaryngologic Clinics of North America 53 (2020) 73--85.
\newblock \href {https://doi.org/10.1016/J.OTC.2019.09.014} {\path{doi:10.1016/J.OTC.2019.09.014}}.

\bibitem{pylsl}
{Python Software Foundation}, \href{https://pandas.pydata.org/}{{Pandas library}} (2023).
\newline\urlprefix\url{https://pandas.pydata.org/}

\bibitem{Pandas}
{W}es {M}c{K}inney, {D}ata {S}tructures for {S}tatistical {C}omputing in {P}ython, in: {S}t\'efan van~der {W}alt, {J}arrod {M}illman (Eds.), {P}roceedings of the 9th {P}ython in {S}cience {C}onference, 2010, pp. 56 -- 61.
\newblock \href {https://doi.org/10.25080/Majora-92bf1922-00a} {\path{doi:10.25080/Majora-92bf1922-00a}}.

\bibitem{MNE}
{Python Software Foundation}, \href{https://pypi.org/project/mne/}{{MNE library}} (2023).
\newline\urlprefix\url{https://pypi.org/project/mne/}

\bibitem{scikit-learn}
F.~Pedregosa, G.~Varoquaux, A.~Gramfort, V.~Michel, B.~Thirion, O.~Grisel, M.~Blondel, P.~Prettenhofer, R.~Weiss, V.~Dubourg, J.~Vanderplas, A.~Passos, D.~Cournapeau, M.~Brucher, M.~Perrot, E.~Duchesnay, Scikit-learn: Machine learning in {P}ython, Journal of Machine Learning Research 12 (2011) 2825--2830.

\bibitem{TensorFlow}
M.~Abadi, A.~Agarwal, P.~Barham, E.~Brevdo, Z.~Chen, C.~Citro, G.~S. Corrado, A.~Davis, J.~Dean, M.~Devin, S.~Ghemawat, I.~Goodfellow, A.~Harp, G.~Irving, M.~Isard, Y.~Jia, R.~Jozefowicz, L.~Kaiser, M.~Kudlur, J.~Levenberg, D.~Man\'{e}, R.~Monga, S.~Moore, D.~Murray, C.~Olah, M.~Schuster, J.~Shlens, B.~Steiner, I.~Sutskever, K.~Talwar, P.~Tucker, V.~Vanhoucke, V.~Vasudevan, F.~Vi\'{e}gas, O.~Vinyals, P.~Warden, M.~Wattenberg, M.~Wicke, Y.~Yu, X.~Zheng, \href{https://www.tensorflow.org/}{{TensorFlow}: Large-scale machine learning on heterogeneous systems}, software available from tensorflow.org (2015).
\newline\urlprefix\url{https://www.tensorflow.org/}

\bibitem{Keras}
F.~Chollet, et~al., Keras, \url{https://keras.io} (2015).

\bibitem{MartinezBeltran:safecar_distractions:2022}
E.~T. {Martínez Beltrán}, M.~{Quiles Pérez}, S.~{López Bernal}, G.~{Martínez Pérez}, A.~{Huertas Celdrán}, Safecar: A brain–computer interface and intelligent framework to detect drivers’ distractions, Expert Systems with Applications 203 (2022) 117402.
\newblock \href {https://doi.org/10.1016/j.eswa.2022.117402} {\path{doi:10.1016/j.eswa.2022.117402}}.

\bibitem{Zhang:poc_drowsiness:2020}
T.~Zhang, H.~Wang, J.~Chen, E.~He, Detecting unfavorable driving states in electroencephalography based on a pca sample entropy feature and multiple classification algorithms, Entropy 2020, Vol. 22, Page 1248 22 (2020) 1248.

\bibitem{Dosovitskiy:CARLA:2017}
A.~Dosovitskiy, G.~Ros, F.~Codevilla, A.~Lopez, V.~Koltun, {CARLA}: {An} open urban driving simulator, in: Proceedings of the 1st Annual Conference on Robot Learning, 2017, pp. 1--16.

\bibitem{Silvera:DReyeVR:2022}
G.~Silvera, A.~Biswas, H.~Admoni, Dreyevr: Democratizing virtual reality driving simulation for behavioural \& interaction research, in: Proceedings of the 2022 ACM/IEEE International Conference on Human-Robot Interaction, 2022, pp. 639--643.

\bibitem{OpenBCI}
{OpenBCI}, \href{https://docs.openbci.com/}{{OpenBCI Documentation}} (2023).
\newline\urlprefix\url{https://docs.openbci.com/}

\bibitem{VarjoAero}
{Varjo}, \href{https://varjo.com/products/aero/\#techspecs}{{Technical Specifications of Varjo Aero}} (2023).
\newline\urlprefix\url{https://varjo.com/products/aero/\#techspecs}

\bibitem{KROM}
{KROM}, \href{https://www.kromgaming.com/en/controllers/k-wheel}{{K-Wheel: Multi-platform gaming wheel}} (2023).
\newline\urlprefix\url{https://www.kromgaming.com/en/controllers/k-wheel}

\bibitem{Klem:10-20system:1999}
G.~H. Klem, H.~L{\"u}ders, H.~H. Jasper, C.~E. Elger, The ten-twenty electrode system of the international federation. the international federation of clinical neurophysiology., Electroencephalography and clinical neurophysiology. Supplement 52 (1999) 3--6.

\bibitem{Hidalgo:drowsiness:2024}
J.~M. Hidalgo~Rogel, E.~T. Mart{\'i}nez~Beltr{\'a}n, M.~Quiles~P{\'e}rez, S.~L{\'o}pez~Bernal, G.~Mart{\'i}nez~P{\'e}rez, A.~Huertas~Celdr{\'a}n, Studying drowsiness detection performance while driving through scalable machine learning models using electroencephalography, Cognitive Computation 16~(3) (2024) 1253--1267.
\newblock \href {https://doi.org/10.1007/s12559-023-10233-5} {\path{doi:10.1007/s12559-023-10233-5}}.

\bibitem{oddball}
Y.-S. Jang, S.-A. Ryu, K.-C. Park, Analysis of p300 related target choice in oddball paradigm, Journal of information and communication convergence engineering 9 (04 2011).
\newblock \href {https://doi.org/10.6109/jicce.2011.9.2.125} {\path{doi:10.6109/jicce.2011.9.2.125}}.

\bibitem{p300wave}
T.~Picton, The p300 wave of the human event-related potential, Journal of clinical neurophysiology : official publication of the American Electroencephalographic Society 9 (1992) 456--79.
\newblock \href {https://doi.org/10.1097/00004691-199210000-00002} {\path{doi:10.1097/00004691-199210000-00002}}.

\bibitem{Rathi:auth_BCI:2021}
N.~Rathi, R.~Singla, S.~Tiwari, A novel approach for designing authentication system using a picture based p300 speller, Cognitive Neurodynamics 15~(5) (2021) 805--824, doi:10.1007/s11571-021-09664-3.

\bibitem{LopezBernal:authentication:2023}
E.~López~Bernal, S.~López~Bernal, G.~Martínez~Pérez, A.~Huertas~Celdrán, Evaluation of data processing and machine learning techniques in p300-based authentication using brain-computer interfaces (2023).
\newblock \href {http://arxiv.org/abs/2311.05270} {\path{arXiv:2311.05270}}.

\bibitem{Lopez_Bernal:cyberattacks_implants:2020}
S.~{L\'opez Bernal}, A.~{Huertas Celdrán}, L.~{Fern\'andez Maim\'o}, M.~T. {Barros}, S.~{Balasubramaniam}, G.~{Mart\'inez P\'erez}, Cyberattacks on miniature brain implants to disrupt spontaneous neural signaling, IEEE Access 8 (2020) 152204--152222.

\bibitem{Lopez_Bernal:cyberBCI:2021}
S.~L\'{o}pez~Bernal, A.~Huertas~Celdr\'{a}n, G.~Mart\'{\i}nez~P\'{e}rez, M.~T. Barros, S.~Balasubramaniam, Security in brain-computer interfaces: State-of-the-art, opportunities, and future challenges, ACM Computing Surveys 54~(1) (Jan. 2021).

\bibitem{Ramadan:controlSignalsReview:2017}
R.~A. Ramadan, A.~V. Vasilakos, {Brain computer interface: control signals review}, Neurocomputing 223 (2017) 26--44.

\bibitem{Dwivedi:metaverse_multidisciplinary:2022}
Y.~K. Dwivedi, L.~Hughes, A.~M. Baabdullah, S.~Ribeiro-Navarrete, M.~Giannakis, M.~M. Al-Debei, D.~Dennehy, B.~Metri, D.~Buhalis, C.~M. Cheung, K.~Conboy, R.~Doyle, R.~Dubey, V.~Dutot, R.~Felix, D.~Goyal, A.~Gustafsson, C.~Hinsch, I.~Jebabli, M.~Janssen, Y.-G. Kim, J.~Kim, S.~Koos, D.~Kreps, N.~Kshetri, V.~Kumar, K.-B. Ooi, S.~Papagiannidis, I.~O. Pappas, A.~Polyviou, S.-M. Park, N.~Pandey, M.~M. Queiroz, R.~Raman, P.~A. Rauschnabel, A.~Shirish, M.~Sigala, K.~Spanaki, G.~{Wei-Han Tan}, M.~K. Tiwari, G.~Viglia, S.~F. Wamba, Metaverse beyond the hype: Multidisciplinary perspectives on emerging challenges, opportunities, and agenda for research, practice and policy, International Journal of Information Management 66 (2022) 102542.

\bibitem{Zhang:BCI_AI:2020}
X.~Zhang, Z.~Ma, H.~Zheng, T.~Li, K.~Chen, X.~Wang, C.~Liu, L.~Xu, X.~Wu, D.~Lin, H.~Lin, The combination of brain-computer interfaces and artificial intelligence: applications and challenges, Annals of Translational Medicine 8 (2020) 712--712.
\newblock \href {https://doi.org/10.21037/ATM.2019.11.109} {\path{doi:10.21037/ATM.2019.11.109}}.

\bibitem{Ju:BCI_FL:2020}
C.~Ju, D.~Gao, R.~Mane, B.~Tan, Y.~Liu, C.~Guan, Federated transfer learning for eeg signal classification, in: 2020 42nd Annual International Conference of the IEEE Engineering in Medicine \& Biology Society (EMBC), 2020, pp. 3040--3045.
\newblock \href {https://doi.org/10.1109/EMBC44109.2020.9175344} {\path{doi:10.1109/EMBC44109.2020.9175344}}.

\bibitem{MartinezBeltran:DFL_survey:2023}
E.~T. Mart{\'i}nez~Beltr{\'a}n, M.~Quiles~P{\'e}rez, P.~M. S{\'a}nchez~S{\'a}nchez, S.~L{\'o}pez~Bernal, G.~Bovet, M.~Gil~P{\'e}rez, G.~Mart{\'i}nez~P{\'e}rez, A.~Huertas~Celdr{\'a}n, {Decentralized Federated Learning: Fundamentals, State of the Art, Frameworks, Trends, and Challenges}, IEEE Communications Surveys \& Tutorials In press~(1-1) (2023) 1--1.
\newblock \href {https://doi.org/10.1109/COMST.2023.3315746} {\path{doi:10.1109/COMST.2023.3315746}}.

\bibitem{Leuthardt:RisksBCI:2021}
E.~C. Leuthardt, D.~W. Moran, T.~R. Mullen, Defining surgical terminology and risk for brain computer interface technologies, Frontiers in Neuroscience 15 (2021).
\newblock \href {https://doi.org/10.3389/fnins.2021.599549} {\path{doi:10.3389/fnins.2021.599549}}.

\bibitem{Biswas:cybersickness:2024}
N.~Biswas, A.~Mukherjee, S.~Bhattacharya, \href{https://doi.org/10.1145/3670008}{“are you feeling sick?” – a systematic literature review of cybersickness in virtual reality}, ACM Comput. Surv. 56~(11) (Jun. 2024).
\newblock \href {https://doi.org/10.1145/3670008} {\path{doi:10.1145/3670008}}.
\newline\urlprefix\url{https://doi.org/10.1145/3670008}

\bibitem{Liu2024}
Y.~Liu, R.~Liu, J.~Ge, Y.~Wang, Advancements in brain-machine interfaces for application in the metaverse, Frontiers in Neuroscience 18 (2024).
\newblock \href {https://doi.org/10.3389/fnins.2024.1383319} {\path{doi:10.3389/fnins.2024.1383319}}.

\bibitem{MartinezBeltran:BCI_noise:2022}
E.~T. Mart{\'i}nez~Beltr{\'a}n, M.~Quiles~P{\'e}rez, S.~L{\'o}pez~Bernal, A.~Huertas~Celdr{\'a}n, G.~Mart{\'i}nez~P{\'e}rez, Noise-based cyberattacks generating fake p300 waves in brain--computer interfaces, Cluster Computing 25~(1) (2022) 33--48.
\newblock \href {https://doi.org/10.1007/s10586-021-03326-z} {\path{doi:10.1007/s10586-021-03326-z}}.

\bibitem{Lopez_Bernal:jamming:2022}
S.~López~Bernal, A.~Huertas~Celdrán, G.~Martínez~Pérez, Neuronal jamming cyberattack over invasive bcis affecting the resolution of tasks requiring visual capabilities, Computers \& Security 112 (2022) 102534.

\bibitem{LopezBernal:taxonomy_attacks:2021}
S.~López~Bernal, A.~Huertas~Celdrán, G.~Martínez~Pérez, Eight reasons to prioritize brain-computer interface cybersecurity, Commun. ACM 66~(4) (2023) 68–78.
\newblock \href {https://doi.org/10.1145/3535509} {\path{doi:10.1145/3535509}}.

\end{thebibliography}

\end{document}